\documentclass[12pt,preprint]{aastex}

\newcommand{\mcmc}        {M_{\rm CMC}}
\newcommand{\ecmc}        {\epsilon_{\rm CMC}}
\newcommand{\tg}          {t_{\rm grow}}
\newcommand{\mdisk}        {M_{\rm d}}
\newcommand{\dt}          {\Delta t}
\newcommand{\fmc}          {f_{mc}}
\newcommand{\mgal}        {M_{\rm galaxy}}

\newcommand{\etal}{\textrm{et~al. }}
\newcommand{\ie}{\textrm{i.e. }}
\newcommand{\eg}{\textrm{e.g. }}

\newcommand{\cf}{\textrm{cf. }}

\shorttitle{Destruction of Bars by CMCs}
\shortauthors{Shen \& Sellwood}

\begin{document}

\title{The Destruction of Bars by Central Mass Concentrations} 

\author{Juntai Shen \& J. A. Sellwood}
\affil{Department of Physics and Astronomy, Rutgers University, 136
Frelinghuysen Road, Piscataway, NJ 08854}
\email{shen@physics.rutgers.edu; sellwood@physics.rutgers.edu}
\slugcomment{\textit{To appear in ``The Astrophysical Journal''}}

\begin{abstract}
More than two thirds of disk galaxies are barred to some degree.  Many
today harbor massive concentrations of gas in their centers, and some
are known to possess supermassive black holes (SMBHs) and their
associated stellar cusps.  Previous theoretical work has suggested
that a bar in a galaxy could be dissolved by the formation of a mass
concentration in the center, although the precise mass and degree of
central concentration required is not well-established.  We report an
extensive study of the effects of central masses on bars in
high-quality $N$-body simulations of galaxies.  We have varied the
growth rate of the central mass, its final mass and degree of
concentration to examine how these factors affect the evolution of the
bar.  Our main conclusions are: (1) Bars are more robust than
previously thought. The central mass has to be as large as several
percent of the disk mass to completely destroy the bar on a short
timescale. (2) For a given mass, dense objects cause the greatest
reduction in bar amplitude, while significantly more diffuse objects
have a lesser effect. (3) The bar amplitude always decreases as the
central mass is grown, and continues to decay thereafter on a
cosmological time-scale. (4) The first phase of bar-weakening is due
to the destruction by the CMC of lower-energy, bar-supporting orbits,
while the second phase is a consequence of secular changes to the
global potential which further diminish the number of bar-supporting
orbits.  We provide detailed phase-space and orbit analysis to support 
this suggestion.

Thus current masses of SMBHs are probably too small, even when dressed
with a stellar cusp, to affect the bar in their host galaxies.  The
molecular gas concentrations found in some barred galaxies are also
too diffuse to affect the amplitude of the bar significantly.  These
findings reconcile the apparent high percentage of barred galaxies with
the presence of central massive concentrations, and have important
implications for the formation and survival of bars in such galaxies.
\end{abstract}


\section{Introduction}
Bars are one of the most prominent morphological features presented by
disk galaxies; Eskridge \etal (2000) find that more than two thirds of
disk galaxies are either strongly or weakly barred in the
near-infrared band.

Central mass concentrations (CMCs) are also frequently found in
galaxies of all types, including barred galaxies. A few examples of
CMCs are:
\begin{itemize}
\item Large condensations of molecular gas with scales of $0.1\sim2$
kpc and masses of $10^7\sim 10^9M_\odot$ are found in the central
regions (e.g. Sakamoto \etal 1999; Regan \etal 2001).  They are
particularly evident in barred galaxies, where they are believed to be
created by bar-driven inflow (\eg Athanassoula 1992; Heller \& Shlosman 1994).

\item Central supermassive black holes (SMBHs) seem to be a ubiquitous
component in spiral galaxies as well as in ellipticals.  Typical
masses are in the range of $10^6\sim 10^9M_\odot$, and there is a
loose correlation with bulge luminosity (Magorrian \etal 1998),
suggesting the mass of the SMBH is roughly $0.006M_{\rm bulge}$, where
$M_{\rm bulge}$ is the mass of the bulge.  Recent measurements with
improved accuracy constrained the typical mass of a SMBH to be about
$10^{-3}M_{\rm bulge}$ (Ferrarese \& Merritt 2000; Gebhardt \etal
2000; Tremaine \etal 2002).  Furthermore, a black hole is thought to
be surrounded by a stellar cusp (Young 1980), which augments the
effective mass of a SMBH for the purpose of large-scale morphological
changes.

\item Dense star clusters are found near the centers of many spiral
galaxies (Carollo 2003; Walcher \etal 2003), which are young, very
compact (a few to up to 20 pc) and relatively massive ($10^6 \sim 10^8
M_\odot$).
\end{itemize}
For the purposes of this paper, a CMC is any sufficiently large mass
at the center of galaxy that is likely to have a dynamical effect on
the evolution of its host galaxy.  In other words we focus on the
dynamical consequences of such objects regardless of their nature.

Studies of single-particle dynamics in rotating bar potentials with a
CMC (Hasan \& Norman 1990; Hasan, Pfenniger \& Norman 1993), and some
limited $N$-body simulations (see below) have given rise to a general
belief that a CMC can weaken or dissolve the bar.  Yet CMCs appear to
coexist with bars in many spiral galaxies.  In order to determine
whether this presents a genuine paradox, we need to know how massive a
CMC is needed to destroy the bar completely and on what timescale.

The best way to address these questions is with fully self-consistent
$N$-body simulations.  Unfortunately, different papers in the
literature give apparently discrepant results.  The simulations by
Norman, Sellwood \& Hasan (1996) showed that a 5\% mass can cause the
bar to dissolve on a dynamical timescale.  That by Friedli (1994),
which included both stars and gas, indicated that objects with 2\% of
the disk mass could dissolve the bar on the time scale of about 1~Gyr.
Hozumi \& Hernquist (1998), who employed a 2-D self-consistent field
(SCF) method, found that black holes with 0.5 to 1\% of the total disk
mass are sufficient to weaken the bar substantially within a few Gyrs.
Berentzen \etal (1998) found, from a single experiment, that the gas
inflow driven by a stellar bar concentrates a gas mass some 1.6\%
of that of the galaxy into the center, which causes the bar to decay
on a timescale of 2 Gyrs.  Bournaud \& Combes (2002) also found that
bars are very fragile. 

There are a number of reasons to regard these results and their
implications as tentative.  First, the high frequency of bars in
galaxies suggests they cannot be destroyed with any great efficiency
(Miller 1996).  Indeed, the massive gas content of the central regions
of some barred galaxies (Sakamoto \etal 1999; Regan \etal 2001), as
large as a few $10^9 M_\odot$ in some cases, suggest that the
destruction of bars by CMCs is less efficient than claimed.

Second, large discrepancies between the different experimental results
suggest possible numerical problems with the simulations or
misunderstood implications.  The simulations to study these processes
are highly challenging numerically: there is a wide range of
timescales on which the particles need to be integrated, and a large
number of particles is also very important in order to make the mass
of an individual particle much smaller than the central mass and to
minimize Poisson noise.

Moreover, none of the previous studies presented a {\em systematic}
exploration of the parameter space relevant to the evolution of bars.
In particular, most restricted themselves to varying the mass of
central objects, and unfortunately did not pay close attention to the
``compactness'' of the CMC, which we show here to be almost as
important.

Studies of bar-forming mechanisms (Sellwood 2000) and secular
evolution of barred galaxies are seriously hampered by our inadequate
understanding of the influence of CMCs on the bar.  The existence of
CMCs in early stages of galaxies makes this problem even more
urgent. There is mounting evidence to indicate that massive black
holes grow very rapidly in their early stages, some reaching $\ga
10^9M_\odot$ at $4\la z \la 6$ (Fan \etal 2001; Vestergaard 2004).

Our main motivation of this work is to verify (or otherwise) previous
results, and to determine the timescale on which the bar is weakened
by a CMC, the critical mass of the CMC which causes rapid bar
dissolution, and other parameters vital to the bar-weakening process
by CMCs.  We have carried out a systematic investigation with
extensive high-quality simulations.  Our aim in this study, is not to
attempt perfect realism, but rather to study the dynamics of bar
destruction in isolation from other possible evolutionary behavior.

\section{Model and Simulation Details}
\label{sec:model}

\subsection{Model setup}
\label{sec:setupmodel}

In outline, our overall strategy is to first construct a model galaxy
which is bar unstable.  We follow its evolution until the bar rotates
at nearly constant pattern speed (\ie rotation rate) and with a
roughly constant amplitude, in the absence of a CMC.  Long after the
bar has settled, we grow a CMC from zero to its final mass within a
certain growth time, and study how the bar is weakened in response to
the CMC growth.  We learn how the bar is affected by some parameter,
such as the growth time of the CMC, from a series of simulations as we
change that parameter while keeping the rest fixed.

\subsubsection{Self-consistent bar models}
\label{sec:disks}
In this work we mainly study two initial bars formed in different
ways, which we name weak and strong initial bars because of their
relative strengths.  We create the weak initial bar simply by evolving
a Kuz'min-Toomre (K-T) disk, which has the surface density
distribution
\begin{equation}
\Sigma(R)= \frac{\mdisk}{2\pi a^2} \left( 1+ \frac{R^2}{a^2}\right) ^{-3/2}.
\end{equation}
Here $R$ is the cylindrical radius and $a$ is a length scale.  We
select disk particles from a distribution function designed to yield
the Toomre stability parameter $Q\simeq 1.5$ (Kalnajs 1976;
Athanassoula \& Sellwood 1986).  We truncate the K-T disk at $R=5a$,
spread the particles vertically as the isothermal sheet with a
locally-defined equilibrium vertical velocity dispersion, and embed it
in the halo (see \S\ref{sec:halo}).  The values of the parameters used
in the initial setup are listed in Table 1.

We allow the model, which is globally unstable, to form a bar.  Some
time after the bar develops, the well-known buckling instability
(Toomre 1966; Raha \etal 1991; Sellwood \& Wilkinson 1993) thickens
the bar. After these events, we obtain a disk containing a long-lived
bar having moderate strength, which we designate our weak initial
bar.

In order to create a strong initial bar, we add fresh particles on
circular orbits in the disk mid-plane to a similar weak bar model,
which started from a slightly thicker initial disk, so that the
buckling instability was milder.  We adopt one of the addition rules
described in Sellwood \& Moore (1999), selecting the angular momentum
$J$ of each new particle from a Gaussian distribution with mean
$\bar{J}=1+0.0025t$ (in the units described in the next paragraph) and
dispersion $\sigma_J=0.5$ (also see Table 1); empirically, we find
these parameters cause the bar to strengthen.  This particle-adding
process mimics the gradual growth of the disk in a simplified way.

We adopt $a$ and $M$ as our units of length and mass, respectively,
and our time units are therefore dynamical times $\sqrt{a^3/GM}$.
From here on all quantities are expressed in units such that $G=M=a=1$
unless otherwise noted.  These units can be scaled to physical values
as desired; we adopt one possible scaling, choosing $M = 5\times
10^{10}M_\odot$ and $a = 3\;$kpc, which implies a unit of time of
roughly $1.2\times 10^7$yr.  

\subsubsection{Central Mass Concentrations}
\label{sec:cmcs}
We mimic a CMC as a Plummer sphere for simplicity, \ie its potential
is given by:
\begin{equation}
\Phi_{\rm CMC}(r)=-\frac{G\mcmc(t)}{\sqrt{r^2 + \ecmc ^2}},
\end{equation}
where $\mcmc(t)$ is the CMC mass.  We regard the scale-length of the
CMC, $\ecmc$ (or softening length), as a physically interesting
parameter, since it controls the compactness of the CMC.  For example,
we use a very small $\ecmc$ ($\sim$ a few pc, much less than the
influence radius of a SMBH with about one percent of galactic mass)
for a ``hard'' CMC to mimic a SMBH and relatively large $\ecmc$
($\sim$ a few hundred pc) for a ``soft'' CMC to represent a massive gas
concentration.

We grow the CMC by increasing its mass according to the relation
\begin{equation}
\mcmc(t)=\left\{
  \begin{array}{ll}
    0 & \tau < 0, \\
    M_{\rm CMC} \sin^2 \left( \pi\tau/2  \right) & 0 \le \tau \le 1,\\
    M_{\rm CMC} & \tau > 1,
  \end{array}\right.  \label{eqn:bhmass} 
\end{equation}
with $\tau=(t-t_{\rm CMC})/\tg$, for a CMC introduced at $t_{\rm
CMC}$.  This sinusoidal time dependence is almost exactly the same as
the cubic-type form adopted by Merritt \& Quinlan (1998), Hozumi \&
Hernquist (1998), and others.  Note that in this work we hold $\ecmc$
fixed as the CMC mass is grown, in contrast to Norman \etal (1996) who
decreased $\ecmc$ at fixed mass.

We adopt $\tg \ga 50$ for most of our simulations.  Even though $\tg$
may not be comparable to the rotational period of our initial bars
($\sim 30$ or 50 time units), we regard this as adiabatic growth
because $\tg$ is much longer than the orbital periods of particles
near the disk center.

\subsubsection{Halo component}
\label{sec:halo}
We include a rigid halo in most of our simulations, having a potential
of the form
\begin{equation}
\label{eqn:halopotential}
\Phi_{\rm halo}= \frac{V_0^2}{2}\ln \left( 1+ \frac{r^2}{c^2} \right).
\end{equation}
This choice yields an asymptotically flat circular velocity of $V_0$
for $r\gg c$, with $c$ being the ``core radius.''  We choose a large
core radius, $c=30a$, and $V_0=0.7(GM/a)^{1/2}$.

We include this large-core halo as a rigid component instead of a
``live'' one represented by real particles, mainly for computational
economy.  In \S\ref{sec:livehalotest} we demonstrate that a live halo,
with very similar potential to the rigid one, gives nearly same evolution
in the bar-destruction process by CMCs.  So we are assured that our
results are not compromised by this simplification.

\subsection{Numerical methods}
\label{sec:code}
In most of our simulations, we employ the 3-D cylindrical polar
grid-based $N$-body code described in detail by Sellwood \& Valluri
(1997).  We solve separately for the three components of the
gravitational acceleration and for the potential of the mass
distribution using Fast Fourier Transforms (FFTs) in the vertical and
azimuthal directions and by direct convolution in the radial
direction.  The gravitational field at a distance $d$ from a unit mass
follows the standard (Plummer sphere) softened potential
$\Phi(d)=-G/\sqrt{d^2+\epsilon^2}$ where the softening length,
$\epsilon$, is a constant. We generally employ more than one million 
particles in our runs; most numerical parameters adopted are
listed in Table 1.

We advance the motion of each particle using simple time-centered
leap-frog in Cartesian coordinates.  A fixed time step for all
particles is impractical because of the wide range of time-scales, and
we vary the time step length in two separate ways.  Our fundamental
time step, $\dt$, defines the interval between successive
determinations of the gravitational field of all the particles.  We
divide the computation volume into 4 zones and advance particles in
the outer zones using longer time steps; the spherical zone boundaries
are at $r = 0.6$, $r = 1.0$, and $r = 2.0$, and time steps are doubled
in each successive zone.  Contributions to the gravitational field
from the outer zones to accelerations near the center are computed
from time-interpolated estimates of the slowly-changing outer mass
distribution (Sellwood 1985).

We also implement ``guard shells'' (see Appendix \ref{app:guard})
around a CMC in which the most rapidly moving particles are integrated
on shorter time steps; the shortest step can be as small as $\dt/2^9$
for hard CMCs.  Since the gravitational field inside the tiny guard
shells is dominated by the CMC, we do not update the field of the
mobile particles during these sub-steps.

For test runs with a live halo, we use a hybrid PM scheme (Appendix B
of Sellwood 2003), in which the self-gravity of the disk is computed
on a high-resolution cylindrical polar grid while that of the halo is
computed using a surface harmonic expansion on a spherical grid.

The code conserves total energy and angular momentum to a satisfactory
degree in these challenging experiments.  In our fiducial run
(\S\ref{sec:fiducial}), $|\Delta E/E|
\approx 0.5\%$ and $|\Delta L_z/L_z| \approx 0.9\%$ over 100 time units,
once CMC growth has ceased.

\subsection{Measure of bar strength}
\label{sec:A_measure}
Our estimator of bar strength is
\begin{equation}
\label{eqn:Adef}
A(t) = \frac{1}{N} \left| \sum _{j=1}^{N}\exp[2i\theta_j]\right|
\end{equation}
which is the relative amplitude of the bisymmetric ($m=2$) Fourier
component of the mass density averaged over certain inner radial range
where the bar dominates.  Here $\theta_j$ is the coordinate of each of
the $N$ particles in the radial range $0.5 < R < 2$ in the simulation
at time $t$.  Our strong initial bar has a greater radial extent, and
therefore $|A|$ underestimates its true amplitude, but we have held
this radial range fixed for both bars in order to minimize differences
between the quantities measured; we are most interested in changes to
$|A|$ and not its absolute value.

We also measure the {\it ellipticity}, $e \equiv 1- b/a$, where $a$
and $b$ are respectively the semimajor and semiminor axes of the bar.
We estimate $e$ by fitting iso-density contours using the Image
Reduction and Analysis Facility (IRAF) to smooth projected density
profiles constructed from the particle distribution using an adaptive
kernel scheme (Silverman 1986).  Observers generally use the maximum
value of ellipticity as the measure of bar amplitude (\eg Laurikainen,
Salo \& Rautiainen 2002).  However, the maximum ellipticity in our
simulations sometimes fluctuates due to complications such as spiral
waves.  We therefore use the ellipticity measured at certain
reasonably-chosen fixed semi-major axis, which yields a measure with
much smaller fluctuations.  Laurikainen, Salo \& Rautiainen (2002)
have shown that the ellipticity of a bar correlates quite well with
the maximum relative tangential force $Q_{\rm b}$ (Buta \& Block
2001), another indicator of bar strength, in the bar region.

We find that the two estimators of bar strength, $A$ and $e$, behave
in similar ways as the CMC grows (see \S\ref{sec:fiducial}).  We
therefore generally rely on $A$ only, since it is much easier to
calculate from an $N$-body simulation.

\section{Results}
\label{sec:results}
We first present the result of a fiducial run in \S\ref{sec:fiducial},
then describe results from our systematic exploration of parameter
space; checks of our simulations are discussed in \S\ref{sec:tests}.

\subsection{A fiducial run}
\label{sec:fiducial}
\begin{figure}[t]
\centerline{\includegraphics[angle=0, width=0.8\hsize]{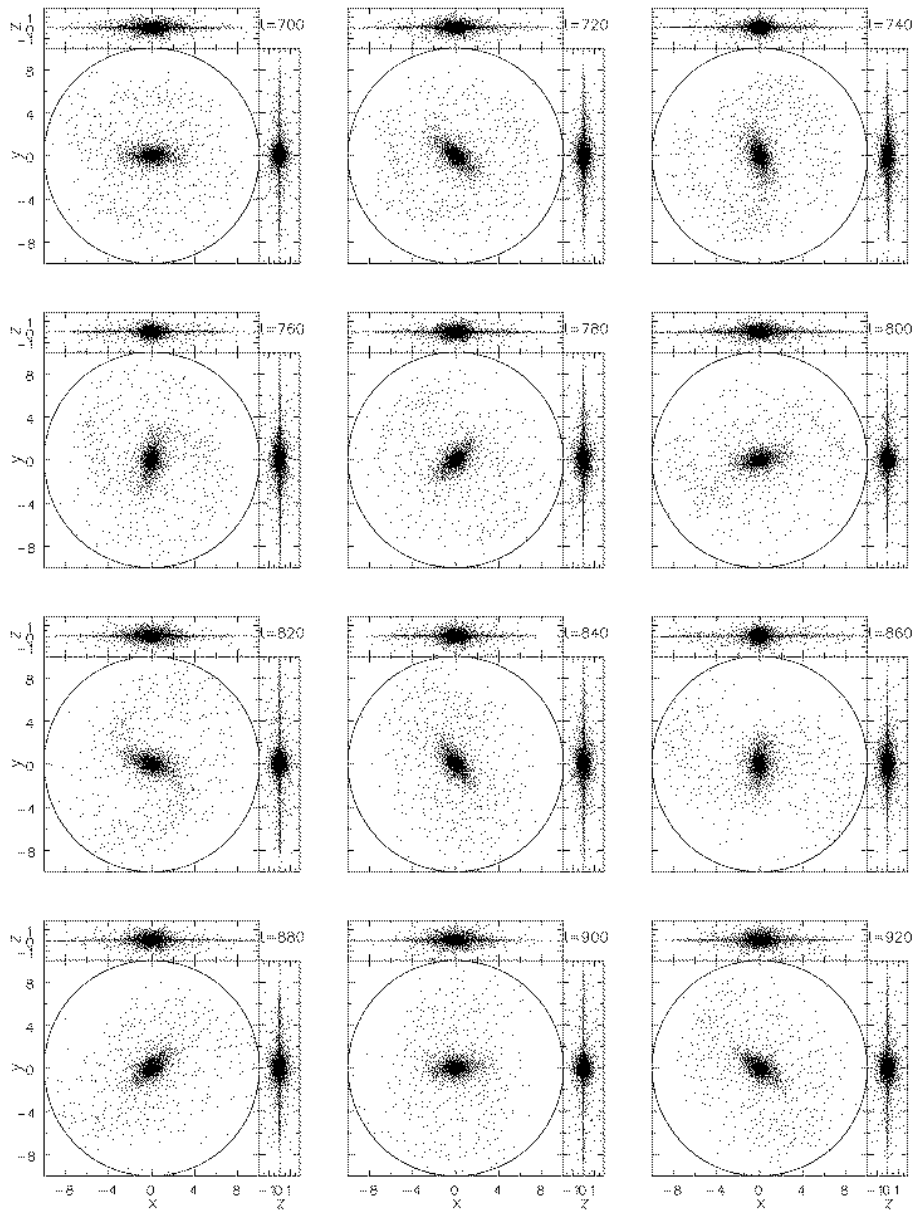}}
\caption{(a) Snapshots of particle positions showing the evolution of
the fiducial model. Only about 1 in 500 particles is shown. The
compact CMC with $\mcmc=0.02$ is grown from $t=700$ to 750 according
to Eq.~(\ref{eqn:bhmass}). The vertical extent of the grid in the
simulation is about twice that shown in the figure.}
\label{fig:snapshots}
\end{figure}

Figure~\ref{fig:snapshots} illustrates how a typical barred model
reacts as a CMC with mass $\mcmc=0.02\mdisk$ and scale-length
$\ecmc=0.001$ is grown according to Eq.~(\ref{eqn:bhmass}) with
$\tg=50$. Figure~\ref{fig:snapshots}(a) shows snapshots of particle
positions and Figure~\ref{fig:snapshots}(b) the corresponding
iso-density contours, obtained by smoothing every particle with an
adaptive kernel.

This experiment starts with the strong initial bar, which rotates
steadily with period $\sim 50$ at constant amplitude.  The CMC starts
to grow at the first time shown ($t=700$) in
Figure~\ref{fig:snapshots}, and its mass rises to $0.02\mdisk$ by
$t=750$.  The bar is weakened by the CMC, but not destroyed, and
retains a moderate strength to the last time shown.

The heavy solid line in Figure~\ref{fig:ell_t} represents $A(t)$, the
evolution of the amplitude defined in (\ref{eqn:Adef}), for our
fiducial run.  The bar weakens rapidly as the CMC mass rises, but the
bar amplitude decreases much more slowly after $t=750$.
Figure~\ref{fig:ell_t} also shows the time evolution of the ellipticity
$e$.  The striking similarity of the overall evolutionary trends of
these two estimators assures us that $A$ is indeed a good bar-strength
indicator.

The scale-length of this CMC is roughly two orders of magnitude
smaller than the influence radius ($r_h=GM_{\rm BH}/\sigma^2 \sim
0.12$; here $\sigma \sim 0.4$ is the flat central value of the velocity
dispersion before the CMC is grown) of a SMBH with the same mass, so
this CMC can be regarded as a SMBH-type object.  However, its mass is of
course much greater than typical SMBH masses in spiral galaxies, which
are about $10^{-3}M_{\rm bulge}$ (Ferrarese \& Merritt 2000; Gebhardt
\etal 2000; Tremaine \etal 2002).  Our intention here is to demonstrate
that a bar can survive even with such an extraordinarily massive hard
CMC.

\subsection{Bar amplitude \mbox{\boldmath $A$} vs. \mbox{\boldmath $\tg$}}
\label{sec:A-tgrowth}
Figure~\ref{fig:A-tgrowth} shows how the bar amplitude evolves for
different $\tg$; the uppermost line indicates that the bar amplitude
stays roughly constant in a comparison run with no central mass.  The
oscillations visible around $t=800$ and $t=1200$ are due to
interference by spiral features just outside the bar region that have
different pattern speeds from that of the bar.  They do not cause any
lasting change to the bar amplitude.

The other lines show that bar is rapidly weakened as the central mass
grows.  Shortly after the CMC mass reaches its maximum value, the bar
amplitude $A$ decays on a much longer timescale for this fixed central
mass.  The transition between the two trends becomes less sharp for
large $\tg$ ($\ga 200$). Thus, we confirm, as found in previous work
(e.g.\ Norman \etal 1996; Hozumi \& Hernquist 1998), that that $\tg$
is not very important in determining the final bar amplitude long
after CMC growth, even when the mass is introduced impulsively and not
adiabatically.  We use $\tg=50$ in all subsequent work.

\subsection{Bar amplitude \mbox{\boldmath $A$} vs. \mbox{\boldmath $\ecmc$}}
\label{sec:A-ecmc}
The importance of the compactness (\ie the scale-length) of CMCs is
explicitly shown in Figure~\ref{fig:A-ecmc}.  The bar amplitude in
Figure~\ref{fig:A-ecmc} is measured at a fixed time, long ($250$ time
units) after the CMC starts to grow.  The final $\mcmc = 0.02\mdisk$
for all runs in Figure~\ref{fig:A-ecmc}.  The solid and dashed curves
represent experiments with weak and strong initial bars, respectively.
Both types of initial bar show a similar trend: {\em hard CMCs are
much more destructive to bars than are softer ones}.  Note that the
trend of bar amplitude with $\ecmc$ is flattening out as the CMC
becomes very hard.

The typical sizes of molecular gas concentrations in the centers of
galaxies range from a few hundred parsecs to 2 kpc or so
(e.g. Sakamoto \etal 1999, Regan \etal 2001), which corresponds to
$0.1\sim 1$ in our simulation units.  Thus diffuse gaseous CMCs cause
much less damage to bars than would equally massive but significantly
more compact counterparts such as SMBHs.

\subsection{Bar amplitude \mbox{\boldmath $A$} vs. \mbox{\boldmath $\mcmc$}}
\label{sec:A-mcmc}
Figure~\ref{fig:A-mcmc} shows the final bar amplitude at the end of
simulations ($\sim 6$ Gyrs after CMC growth) as a function of $\mcmc$,
for both types of initial bar.  This figure indicates that the final
bar amplitude decreases continuously as $\mcmc$ is increased, and
again we see that hard CMCs cause much more damage to bars than do
soft ones (also see Figure~\ref{fig:A-ecmc}).  For both initial bars,
we find that the bar is effectively destroyed when the mass of a hard
($\ecmc=0.001$) CMC is greater than 4\% or 5\% of $\mdisk$, whereas
soft ($\ecmc=0.1$) CMCs as massive as $\sim 0.1\mdisk$ do not destroy
the bar within $\sim 6$ Gyr (500 time units).

\section{Parameter Tests and Numerical Checks}
\label{sec:tests}
These simulations are technically very challenging.  Here we report a
few of our many tests to check that our main results and conclusions
are reliable.  Some of our findings might account for the discrepant
results obtained by other groups. 

\subsection{Numerical parameter tests}
\label{sec:basictests}
Figure~\ref{fig:basictests} presents tests (for the weak initial bar)
of the number of particles $N$, grid size and particle softening,
respectively.  We have conducted a similar suite of tests for the
strong initial bar, with similar results.  The amplitude difference
shown in the top panel for the two experiments with different $N$
starts and remains very small (it is hard to achieve precisely the
same bar parameters).  There is a mild dependence on particle
softening, slightly more pronounced for the strong initial bar, which
we discuss below.  These figures confirm that variations of particle
number and grid size by a factor of two or more around our adopted
values make negligible difference to our main results.

\subsection{Time step}
\label{sec:timesteptest}
We do, however, find that the evolution is sensitive to the adopted
time step $\dt$, and one obtains more rapid, but erroneous, bar
destruction by CMCs if the orbit integration is not handled with
sufficient care.

If a single time step is used over the whole simulation domain, $\dt$
needs to be as small as $10^{-5} \sim 10^{-4}$ to ensure that all
particles are integrated accurately (see Appendix \ref{app:guard}),
for a typical simulation containing a hard central mass with
$\mcmc=0.02\mdisk$ and $\ecmc=0.001$.  Time steps this short are, of
course, too expensive for practical use, which is the reason for the
multiple zones and guard shell schemes described above.

Figure~\ref{fig:timesteptest} shows the effect of changing the time
step while all other parameters are held fixed.  When no guard shells
are employed, the bar amplitude at a fixed time becomes weaker as the
time step is increased; there is no indication of convergence even for
very small $\dt$.  The bar amplitude is always greater when guard
shells are employed, and is less sensitive to the particular choice of
the basic $\dt$, since we divide it even more finely as $\dt$ is
increased.  Even with guard shells, the final amplitude does increase
mildly as the time step is reduced.

Since the severity of the time-step problem is greater for harder and
more massive CMCs, it requires great care to determine the threshold
for bar destruction.

\subsection{Live halo tests}
\label{sec:livehalotest}
Figure~\ref{fig:livehalotest} shows, for one case, that the bar
amplitude evolution is hardly affected when the rigid halo is replaced
by a similar live pseudo-isothermal halo with a large core resembling
our rigid form (Eq.~\ref{eqn:halopotential}).  Thus, our
approximation of a rigid halo appears to be adequate.

While not a parameter test, we report here that a {\em dense} or {\em
cuspy} live halo stimulates the growth of a bar, as noted by
Debattista \& Sellwood (2000) and discussed in more detail by
Athanassoula \& Misiriotis (2002) and Athanassoula (2003).  Thus, a
bar in a denser or cuspy halo is {\em even harder} to destroy, which
underscores one of our main conclusions that bars are very robust.

\subsection{Relation to other work}
We are unable to understand all discrepancies with other experiments
in the literature because many authors give insufficient
information.  We believe the parameters adopted by Norman \etal (1996)
were adequate, and their results are largely consistent with ours.

We suspect inadequately short time steps could be a factor in
accounting for the rapid bar decay claimed by some groups.  For
example, the shortest time step (0.4 Myr) used in Bournaud \& Combes
(2002) is equivalent to $\sim 0.033$ in our units, which our
Figure~\ref{fig:timesteptest} suggests is dangerously long.

But an inadequate time step may not be able to account for all the
discrepancies.  For example, Hozumi \& Hernquist (1998) used a timestep
$\dt=0.005$ which should be adequate for $\ecmc=0.01$, but they found
the bar to be quite fragile, inconsistent with our findings here.  We
speculate that the difference might be due to the SCF (self-consistent
field) simulation method they used, in which the gravitational field is
computed from a rather low-order expansion in some set of basis
functions.  The method therefore raises a severe restriction on the
complexity of shapes which their models could support. In particular
such simulations may be unable to sustain a shape which is axisymmetric
near the center while remaining non-axisymmetric in the outer parts,
which might spuriously hasten the decay of bars. 

\section{The Mechanism of Bar Dissolution by a CMC}
\label{sec:mechanism}
A na\"\i ve explanation for the robustness of a bar against the
central mass is that the orbits of most bar-supporting particles are
little affected by the introduction of a light, dense central mass.
This is because most bar-supporting orbits are loops that naturally tend to
avoid the center of the bar due to the Coriolis force in a rotating
frame (Sparke \& Sellwood 1987; Hasan \& Norman 1990; Sellwood \& 
Wilkinson 1993; etc.).  Contopoulos has designated
this main family of bar-supporting orbits as $x_1$ orbits (see
Contopoulos \& Grosb\o l 1989 for a review), but they are also
sometimes described as B-orbits.

A more careful examination of Figure~\ref{fig:A-tgrowth} reveals an
intriguing phenomenon: the bar is generally weakened in two phases,
divided by the time at which the CMC mass growth stops, which are
easily distinguishable unless $\tg$ is extremely long ($< 200$).  The
first phase, which coincides with the CMC mass-growing period, is
clearly controlled by $\tg$, except when it is very abrupt. Thus the
response is effectively instantaneous because typical dynamical times of
particles near the CMC are short in comparison with $\tg$.  The second
phase of bar weakening is on a much longer time scale ($\ga 0.5$
Hubble time) and begins right after the period of CMC mass growth.
Why is the bar weakened by the CMC in two distinct rates?

We suggest that the first phase of bar weakening is due mainly to the
rapid scattering of $x_1$ particles with lower energies.\footnote{In a
potential which rotates steadily at the rate $\Omega_p$, energy is not
conserved, but the Jacobi constant $E_J=E-\mathbf{\Omega_p} \cdot
\mathbf{L}= \frac{1}{2}|\mathbf{\dot{r}}|^2 + \left(\Phi - \frac{1}{2}
\Omega_p^2 R^2 \right)$ is constant along on an orbit (Binney \&
Tremaine 1987).  In this paper we loosely refer to the Jacobi constant
$E_J$ as the energy.}  To support this suggestion, we present a study
of orbits in the mid-plane ($z=0$) of the bar.

One of the more powerful tools of orbit analysis is the surface of
section (SOS) which can be constructed at any accessible energy.  We
adopt Cartesian coordinates in the rotating frame of the bar, with the
major axis of the bar aligned with the $x$-axis, and integrate the
orbits of test particles in a steadily rotating bar potential from
some moment in the simulation.  A SOS records the values of $(y,
\dot{y})$ for a number of orbits all having the same energy, each time
they cross the minor ($y-$) axis with $\dot{x} < 0$.  Particles in
prograde motion in the frame of the bar have $y>0$ when $\dot{x} < 0$,
and particles in retrograde motion have $y<0$ when $\dot{x} < 0$.  An
invariant curve, a series of points all lying on a closed curve in the
SOS, is produced by a regular orbit that conserves a second integral
in addition to the energy.  Irregular, or chaotic, orbits conserve
only one integral ($E_J$) and fill an area in this plane.  The adopted
energy of the SOS determines the maximum possible excursion, $y_{\rm
max}$, on the bar minor-axis, which is the parameter we prefer to
specify.

Figure~\ref{fig:sos_examples} shows the SOS at two energies for the
potential of the rotating bar at $t=700$ before the CMC is grown.  The
series of nested invariant curves on the left arises from the
retrograde ($x_4$) orbit family, which is not very eccentric even in
this strongly non-axisymmetric potential.  Because of their round
shape, $x_4$ orbits do not support the bar and a self-consistent bar
can contain rather little mass moving on orbits of this family.

The other series of nested invariant curves on the right of
Figure~\ref{fig:sos_examples} is of much greater importance.  These
regular prograde orbits are members of the $x_1$ family, and are
widely believed to form the backbone of the bar.
Figure~\ref{fig:x1_examples}(a) shows three examples of $x_1$ orbits
selected from Figure~\ref{fig:sos_examples}(a), showing clearly that
many particles on lower-energy $x_1$ orbits pass very close to the
center; such orbits will be strongly affected by the introduction of
even a low-mass, but dense, CMC.  Regular $x_1$ orbits selected from
the higher-energy SOS (Figure~\ref{fig:sos_examples}b), on the other
hand, are larger (Figure~\ref{fig:x1_examples}b) and generally avoid
the central region.

\subsection{CMC growth phase}
Figure~\ref{fig:sos_evo} shows how the SOS at lower energies changes
as the CMC in our fiducial run grows, corresponding to the first phase
of bar weakening.\footnote{It should be emphasized that these Figures
show the orbit structure in an assumed steady potential taken from the
simulations at a number of different times.  In reality, the gradual
growth of the CMC causes a continuously changing potential.}  The
breakdown of the large regular areas once occupied by the $x_1$ family
is very striking.  More and more bar-supporting particles become
chaotic, as the growing central mass affects orbits that pass at ever
increasing distances.  Since chaotic orbits tend to have more nearly
round long-time average shapes, it is clear that the bar is weakened
in this phase because many regular bar-supporting orbits become
chaotic.

Since the orbital periods of particles near the center are much
shorter than $\tg$, low-energy orbits adjust essentially instantly to
the new potential and the rate of change of bar amplitude tracks the
CMC growth.

\subsection{Structural adjustment}
As the orbits of particles change, the global potential of the system
evolves secularly, an effect that continues even after the CMC has
reached its final mass.  This structural adjustment is illustrated in
Figure~\ref{fig:ymax1.4sos_evo} for the fiducial run ($\ecmc=0.001$,
$\mcmc = 0.02\mdisk$), which shows the slow evolution after $\tg$.

The SOS at $y_{\rm max}=1.4$ keeps evolving even after the CMC has
reached full mass at $t=750$.  The Figure clearly shows that the area
occupied by regular $x_1$ orbits at this lower-energy gradually
diminishes and the $x_2$ orbit family, which does not support bar,
becomes increasingly important among the prograde orbits at this
energy level.  The potential of the system also tends to favor $x_1$
orbits that are somewhat fatter in shape (Figure~\ref{fig:fatter_x1})
at higher energies, since the once-strong bar has now been weakened.

Based on Figure~\ref{fig:A-tgrowth}, this structural adjustment
happens on a very long timescale, $\ga 0.5$ Hubble time.  However,
there is no reason to expect that the rate of this second-phase
bar-weakening should remain constant over long time, in fact it
becomes slightly slower during the course of our simulation
(Figure~\ref{fig:A-tgrowth}).  We also notice that the second phase
bar decay rate is somewhat slower for the weak initial bar than for
the strong initial bar, suggesting that the rate of secular structural
change differs from case to case.

Note that this second phase of bar weakening could not be observed in
studies of single particle dynamics in a rigid potential.  It happens
because the global potential evolves {\it self-consistently}, which
requires an $N$-body simulation.  Furthermore, we find that the area
occupied by regular $x_1$ orbits shrinks rapidly for $\mcmc \ga 0.04
\mdisk$ when the potential is allowed to evolve, which could be
achieved only by further increases in $\mcmc$ if the bar strength is
fixed (also see Hasan \& Norman 1990).

\subsection{Discussion of dissolution mechanisms}

We find that two-phase bar destruction occurs only for compact CMCs of
modest mass ($\la 0.03\mdisk$). A more massive CMC causes a larger
fraction of $x_1$ orbits to become chaotic in the first phase, causing
the second-phase structural adjustment to also happen very
rapidly. Thus the self-consistent bar can no longer withstand the
effect of the massive CMC and starts to dissolve.  This idea may
explain why the critical mass is about the same for both strong and
weak initial bars.

The fact that harder (more compact) CMCs cause more damage to bars
than do softer (more diffuse) ones can also be explained qualitatively
in our picture.  The weaker forces from a soft CMC might be expected
to introduce less chaos.  This expectation is suported by the evidence
in Figure~\ref{fig:sos_cmc_softening}, which compares the SOS for
orbits of similar size in the presence of hard and soft CMCs.
Figure~\ref{fig:sos_cmc_softening}(a) shows the SOS at $y_{\rm
max}=1.2$ at $t=800$ for the fiducial run with a hard ($\ecmc =
0.001$) CMC, while Figure~\ref{fig:sos_cmc_softening}(b) shows the
same for a very diffuse ($\ecmc = 0.1$) CMC. The area of $x_1$ orbit
family in Figure~\ref{fig:sos_cmc_softening}(b) is much larger than
the corresponding region in Figure~\ref{fig:sos_cmc_softening}(a);
similar SOS plots at other energies also show the same result.  This
is in agreement with the finding by Gerhard \& Binney (1985) that the
chaotic region in the SOS increases as the central mass is hardened.

Our picture may also be able to account for the erroneously faster bar
decay resulting from inaccurate orbit integration (\cf
\S\ref{sec:timesteptest}).  If a bar-supporting $x_1$ orbit is not
followed very accurately as it approaches a CMC, values of integrals
may change for numerical reasons, and a particle on a regular $x_1$
orbit could become chaotic.  The enhanced loss of $x_1$ particles will
lead to structural adjustment, making the inverse of this change, a
chaotic orbit converting to a regular one, less likely.  Thus such
errors can only accelerate bar decay.

Figure~\ref{fig:A-tgrowth} also shows that the bar amplitude at the
end of CMC growth does not have the same value for the different
$\tg$, implying that secular structural change may have started the
moment the CMC starts to grow.

Figure~\ref{fig:basictests}(c) showed a mild dependence of the final bar
amplitude on particle softening, indicating that the decay rate in the
second-phase is very slightly accelerated as softening is decreased.  A
slight increase in the rate of collisional relaxation is the most
natural explanation but we stress that the effect is very minor even
though the softening length changes by a factor of five.  Such a weak
dependence indicates that the models are indeed almost collisionless.

\subsection{Is 2-D SOS an adequate tool for 3-D potentials?}
Since 3-D orbit dynamics is complicated and very difficult to
visualize, we do not wish to give a detailed discussion here.  The
preceding discussion concerned only those orbits confined to the $z=0$
plane of the 3-D potential.  This could be misleading, since many cases
are known (\eg Pfenniger 1984) where 2-D orbits are irregular while
3-D orbits, with projected shapes very similar to those in 2-D, may be
regular.  We therefore need to show that allowing 3-D motion does not
change our arguments, especially in regard to the importance of chaos.

The SOS in Figure~\ref{fig:sos_evo}(a), with no CMC, is a nice example
of a strong bar that is supported by a large contribution from regular
2-D $x_1$ orbits, yet it was constructed from a 3-D bar.  It therefore
seems likely that the main bar-supporting orbits are not drastically
altered when 3-D motion is allowed.  Pfenniger \& Friedli (1991) show
that $x_1$-like orbits in 3-D are generally 2:2:1 resonant, that is
they bob up and down about the mid-plane twice for each turn around
the center in the rotating frame of the bar.  Skokos, Patsis \&
Athanassoula (2002a,b) give a more comprehensive discussion.  Since
these orbits are simple 3-D generalizations of the $x_1$ orbit family
that sustains 2-D bars, we expect the essentials of the phase space
structure we find to carry over to 3-D bars.

We also need to show, however, that the evidence for chaos in our 2-D
SOS is in fact accompanied by chaos in the full 3-D potential.  In
order to address this issue, and to complement our 2-D SOS study, we
have carried out a Floquet analysis (Binney \& Tremaine 1987).  The
existence of chaos is indicated by a positive Lyapunov exponent, which
is defined as the $e$-folding divergence rate of two orbits starting
with an infinitesimal separation in phase space.  However, Valluri \&
Merritt (2000) show that Lyapunov exponents in stellar systems are not
good predictors of regular or chaotic orbits, and suggest that a more
practical criterion is whether the exponential divergence {\it
continues} until the phase-space separations reach the system size.

As an example, Figure~\ref{fig:lya_ex} shows the evolution of
phase-space separation from a very small value for three orbits;
the left-hand panel (a) shows the case when motion continues to be
confined to the $z=0$ plane, while the right-hand panel (b) shows cases
in which full 3-D motion is allowed.  The latter orbits start with the
same $(x,y,z=0)$ position, but move away from the mid-plane because
the potential is not perfectly reflection symmetric.  The two orbits
in each panel for which the separation quickly reaches the system size
are clearly chaotic while the orbit for which the separation remains
at some value much smaller than the system size for a long time
($\sim5\;$Gyr) is regular, according to the criterion proposed by
Valluri \& Merritt (2000).  Thus for these three examples, at least,
the (ir)regular nature is unchanged when 3-D motion is permitted.

In order to know whether a region that appears to be chaotic in 2-D
remains fully, or mostly, chaotic when 3-D motion is allowed, we
examine the phase-space separation of many pairs of orbits started at
random in the region in question.  We focus on two 3-D potentials at
$t=740$ and $t=1200$, since both show large areas of chaos at some
energy in the 2-D SOS plots (\ie Figure~\ref{fig:sos_evo}(e) for
$E_J=-0.6385$ at $t=740$ and Figure~\ref{fig:ymax1.4sos_evo}(f) for
$E_J=-0.5164$ at $t=1200$).  For 500 orbits, we choose the initial $y$
at random from the range $-0.15 < y < 0.6$ for $t=740$ and $-0.1 < y <
0.5$ for $t=1200$, the initial $x=0$, and the initial $z$ is one of
five values between $z=\pm 0.2$.  For velocities, we select $V_x$,
$V_y$ and $V_z$ randomly to satisfy the selected $E_J$.  From the
criterion for a chaotic orbit described above
(Figure~\ref{fig:lya_ex}), we find that about 90\% of our
randomly-chosen orbits are chaotic for $E_J=-0.6385$ at $t=740$ and
about 80\% are chaotic for $E_J=-0.5164$ at $t=1200$.

Thus this 3-D test confirms that most of the realistic 3-D orbits are
chaotic, consistent with expectation of the 2-D SOS results.  We
therefore contend that a 2-D SOS indeed provides a simplified, yet
useful, way to picture the underlying 3-D orbital structure in our
study. 

\section{Discussion}
\label{sec:discuss}
The experiments described here are rather artificial.  We simply grow
central masses in a pre-existing bar in order to learn how the bar
is affected by CMCs having a range of properties.  Our intention was to
isolate and understand a single aspect of the dynamics, with as few
complications as possible.  Despite this, and other limitations, it
may be reasonable to apply our main results to real galaxies.

\subsection{Limitations}
We have restricted our simulations to collisionless stellar systems
only, and have not attempted to include gas.  We justify this mainly
because simulations with gas are much more complicated and we have
shown that it requires great care to get even the simpler
collisionless results right.  The forced inflow of gas is of
considerable importance, and star formation, feedback, etc., will also
affect the accumulation of gas near the center.  However, gas mass
fractions in galaxies today are thought to be modest, and the mass
accumulations in the centers of bars appear not to be large enough to
destroy the bars (as we discuss next); thus our approximation of
treating all mobile material as if it were collisionless may capture
most of the essential physics relevant to bar survival.

We have not included a bulge component in our models, largely because it
would increase the number of parameters; we leave a study that includes
realistic bulges for future work.  Our results therefore apply most
directly to small-bulge galaxies (late-type spirals).  While SMBHs are
associated directly with bulges, and not with disks, other types of CMCs
are found in late-type galaxies.  However, we do not expect the bar
dissolution behavior to be substantially affected by the introduction of
a modest bulge component because bulges are relatively diffuse objects.
Even if the bulge density profile has an inner cusp of slope
$r^{-\alpha}$, the enclosed mass rises as $r^{3-\alpha}$ from the center,
and therefore only a tiny fraction of the bulge mass will behave as a hard
CMC.

We have studied two bars with different strengths and found consistent
results.  But the exact properties of a CMC to dissolve a bar may also
depend on other details of the initial bar in host galaxies.  For
example, one might expect long slender bars to be more fragile than
short stubby bars (also see Hasan \& Norman 1990).  The pattern speed
might also affect the bar dissolution process, since Teuben \& Sanders
(1985) showed that dynamically slow bars are supported by different
orbit families than are fast bars.  However, it is hard to vary the
bar pattern speed only while keeping other properties of the bar
unchanged in a self-consistent study.  In fact, the pattern speed and
length of the bar appear to be closely related (Aguerri, Debattista \& 
Corsini 2003).

It may be more appropriate to express the critical CMC mass as a
fraction of the bar mass ($M_{\rm bar}$), since the outer disk
presumably has little effect on the dynamics near the center of a bar.
Unfortunately we found this impractical, as an accurate measure of
$M_{\rm bar}$ alone requires us to separate the bar from the other
components, which is hard to do -- both for simulations and for real
galaxies.

\subsection{Implications}
We have shown that the timescale for bar destruction depends on the
mass of the CMC and that SMBH-like hard CMCs cause much more damage to
bars than do soft ones.  The critical SMBH mass to destroy bars on
short timescale is about 4\% or 5\% of the disk mass for the bars used
in our study.  Ferrarese \& Merritt (2000), Gebhardt \etal (2000), and
Tremaine \etal (2002) estimated the mass of SMBHs found in many
galaxies to be $\sim 10^{-3}$ of the {\em bulge} mass, and bulges are
generally less luminous (\eg Palunas \& Williams 2000), and probably
less massive, than the disk, except for rare cases.  Thus SMBHs in
galaxies are much less than this critical value and are therefore too
small to pose a significant threat to the survival of bars today.  For
the same reason, central dense star clusters in some galaxies (Carollo
2003; Walcher \etal 2003) are also incapable of dissolving bars.

Our finding that bars are more robust than previously thought
eliminates any possible paradox presented by the coexistence of SMBHs
and gas concentrations in the centers of barred galaxies.  It is
widely believed that bars drive gas into the central regions (\eg
Athanassoula 1992), yet our findings indicate they should be able to
survive such massive gas accumulations.  Sakamoto \etal (1999) and
Regan \etal (2001) detected much greater concentrations of molecular
gas in barred galaxies than in unbarred ones.  This is the opposite of
what would be expected from rapid bar destruction by central gas
concentrations, since greater accumulations would be observed in
unbarred ({\em but previously barred\/}) galaxies, provided the gas
concentration survives for some time.

Das \etal (2003) find an inverse correlation between the deprojected
bar ellipticity ($e$) and the ratio of the dynamical mass within the
bulge to that within the bar radius ($\fmc$) in a small sample of 13
nearby barred galaxies taken from the BIMA SONG survey.  They argue
that $\fmc$ is a good indicator, although not a direct measure, of the
central mass concentration, and they further suggest that the
correlation with $e$ may be the consequence of the bar dissolution by
CMCs.

Their correlation needs to be confirmed in a larger sample.  In
particular, 11 of the 13 sample members are in the range of $0.3 < e <
0.62$, while only two have small ellipticities and essentially set the
``tightness'' of their inverse correlation.  If we regard the two
small-$e$ galaxies as unreliable, because of uncertainties in
deprojection (Barnes \& Sellwood 2003), the inverse correlation
between $e$ and $\fmc$ is rather loose.

If indeed real, the correlation would provide an important constraint
on the bar weakening process by CMCs.  For example, the modest diffuse
CMCs they report can not cause {\em rapid} bar destruction ($\la$ a
couple of Gyrs), otherwise most bars should have been destroyed
already.  To avoid this conclusion, one would require bars to be
rejuvenated in a special way, \eg\ strong bars must reform in galaxies
with smaller CMCs and vice versa.  On the other hand, the correlation
does not rule out our finding of slow bar dissolution; bars are
weakened to differing extents, more by a large CMC and less by a small
counterpart, and then survive at their weakened amplitude for a long
time.  Detailed predictions are impossible, however, because we
currently lack a coherent picture of when, and in what sequence, bars
and CMCs formed.

Our results suggest it is unlikely that large fractions of bulges are
formed by bar dissolution, because the CMCs observed in galaxies today
do not destroy bars.  However, the bar weakening process does add some
disk material to bulges, since those orbits which become chaotic because
of the presence of the CMC fill a more nearly spherical volume near the
center of the bar. Chaotic orbits which were quite elongated before the
introduction of the CMC will contribute to the bulge density to a radius
several times that of the usual sphere of influence of a SMBH.
Furthermore, such a process might be more important in young galaxies
when disks/bars were smaller and CMCs could have been proportionately
more massive.

CMCs do not, of course, offer the only means to destroy or weaken
bars.  In fact, we have shown that neither realistic SMBHs nor diffuse
molecular gas clouds can dissolve bars with any great efficiency.
There are other ways to weaken and destroy bars more efficiently than
by growing a CMC.  Bars can be destroyed in minor mergers (Berentzen \etal 
2003 and references therein).  Also
Sellwood \& Moore (1999) found that ongoing spiral activity in the
outer disk triggered by newly formed ``cold'' material accreted onto
the disk, which is not included in our present simulations, could
either complete the destruction of the bar or cause it to grow again;
buckling also weakens bars (\eg Raha \etal 1991).

Our problem is similar to the question of whether a SMBH destroys the
large-scale triaxiality of a typical elliptical galaxy.  As these
objects, which are believed to lack significant figure rotation, are
supported mostly by box orbits and not by tubes, as in our case, the dynamics
is quite different, yet it is interesting that the survival of
triaxiality in this case is also not yet settled.  Holley-Bockelmann
\etal (2002) found that triaxiality may be long-lived even with a SMBH
as massive as $0.01\mgal$, while Merritt \& Quinlan (1998) found that
SMBHs with $0.003\mgal$ are able to destroy large-scale triaxiality
within a Hubble time. Further investigations in this area would also
seem to be warranted.

\section{Conclusions}
\label{sec:conclusions}
We have conducted a systematic study of the effects of central massive
concentrations (CMCs) on bars using high-quality $N$-body simulations.
We have experimented with both strong and weak initial bars and a wide
range of physical parameters of CMCs, such as the final mass,
scale-length and mass growth time $\tg$ etc.  We have demonstrated
that our main findings are insensitive to most numerical parameters in
our simulation, and shown that the time step requires special care 
particularly for the cases with compact CMCs.

We find that, for a given mass, compact CMCs (such as supermassive
black holes) are more destructive to bars than are more diffuse ones
(such as molecular gas clouds in many galactic centers).  We have
shown that the former are more efficient scatterers of bar-supporting
$x_1$ orbits that pass close to the center, therefore decrease the
number of regular $x_1$ orbits and increase the size of the chaotic
region in phase space.

A bar is generally weakened by a CMC in two connected phases.  The bar
strength decreases rapidly as the central mass grows due to the rapid
scattering of stars on lower-energy bar-supporting $x_1$ orbits as
they pass close to the CMC.  The time scale for this first phase is
therefore the orbital period of the stars in the bar or the growth
rate of the central mass, whichever is the longer.  The bar continues
to decay thereafter on a cosmological timescale (\eg $\ga 0.5$ Hubble
time for a compact CMC with $0.02 \mdisk$).  The second phase reflects
slow evolution of the gravitational potential causing a gradual loss
of bar-supporting $x_1$ orbits.

Bars are more robust against CMCs than previously thought: the central
object, even for the most destructive SMBH-like CMCs, has to be as
massive as a few percent of the disk mass to destroy a bar completely
within a Hubble time.  On the other hand, diffuse CMCs need a
tremendous amount of mass ($> 0.1 \mdisk$) to achieve the same
effect.

Our findings clearly show that neither typical SMBHs in spirals
($\mcmc\sim 10^{-3}M_{\rm Bulge}$) nor typical central molecular gas
concentrations (mass $\mcmc \la$ a few percent of $\mdisk$, scale $R
\sim$ a few hundred pcs) can have any significant weakening effect on
the bar within a Hubble time -- the former are generally not
massive enough, whereas the latter are too diffuse.  Thus, our results
can naturally account for the coexistence of CMCs and bars in many
spiral galaxies.


\acknowledgements
We thank the anonymous referee for a thoughtful report and for
suggesting further examination of the 3-D ``chaos'' question.  This work
was supported by NSF grant AST-0098282.


\appendix

\section{Guard shells scheme}
\label{app:guard}
We have devised the following ``guard shells'' scheme in order to
improve the accuracy of orbit integration when particles experience
strong acceleration near a compact central mass.  We divide the volume
around the CMC into many concentric shells and successively halve the
time step for particles in each shell as they approach the CMC
(Figure~\ref{fig:GA}).  Since accelerations in this small region are
dominated by the field of the CMC, we integrate the orbit for these
sub-timesteps in a fixed field, and update the self-consistent part
from the bar and the disk at the basic time step interval $\dt$.

In the region dominated by the CMC, the orbital period $\tau$ of a
test particle moving on a circular orbit satisfies $$
|F_r|= \frac{d\Phi_{\rm CMC}}{dr}=\frac{r \,G \mcmc
}{(r^2+\ecmc^2)^{3/2}} =\left(\frac{2\pi}{\tau} \right)^2 r,
$$ so $$
\tau(r)=2\pi\sqrt{\frac{(r^2+\ecmc^2)^{3/2}}{G \mcmc}}. $$

We choose the outermost guard radius $r_{\rm max}$ so that
$\tau(r_1\!=\!r_{\rm max})/\dt$ is $\ga 100$.  Particles inside
$r_{\rm max}$, which have shorter periods, require finer time steps.
We adopt time steps in adjacent shells that differ by a factor of two,
making it appropriate to require $\tau(r_{i+1}) = \tau(r_i)/2$.  Since
$\tau \propto r^{3/2}$ (if $r \gg \ecmc$), the ratio of adjacent guard
radii is $$ 
\frac{r_{i+1}}{r_i}=\left[ \frac{\tau(r_{i+1})}{\tau(r_i)}
\right]^{2/3}= \left( \frac{1}{2} \right)^{2/3} \approx 0.63.
$$ We keep this ratio constant for simplicity, although this is somewhat
over-conservative for $r \sim \ecmc$.

The shortest possible orbital period is $$
\tau_{\rm min}=\tau(r=0)=2\pi \frac{\ecmc^{3/2}}{\sqrt{G\mcmc}}.
$$ The number of guard shells ($n$), or the boundary of the innermost
zone ($r_{\rm min})$, is determined so that $\tau_{\rm min}/(\dt/2^n)
\ga 50$ say, which we find to be adequate.


\clearpage
\vfill\eject

\begin{table}[t]
\setlength{\baselineskip}{12pt}
\begin{center}
\begin{tabular}{lcc}
   \multicolumn{3}{c}{Table 1.    Summary of the model setup} \\ \hline \hline
   \multicolumn{1}{c}{Quantity} &
   \multicolumn{1}{c}{Strong initial bar}  &
   \multicolumn{1}{c}{Weak initial bar\tablenotemark{a}}  \\ \hline

   \multicolumn{3}{c}{Numerical Parameters} \\ \hline
   Initial number of particles \dotfill & $2.0 \times 10^6$ & $1.2 \times 10^6$ \\
   Final number of particles \dotfill & $2.8 \times 10^6$ \\
   Grid size $(R,\phi,z)$ \dotfill & $55\times64\times375$  &
 $55\times64\times375$  \\
   Vertical plane spacing \dotfill & $0.02$ & $0.02$ \\
   Grid boundaries $(R,z)$ \dotfill & $(20.0,\pm3.74)$ & $(20.0,\pm3.74)$ \\
   Particle softening length \dotfill & $0.02$ & $0.05$ \\
   Time step $\dt_0$\ --- w/o CMC \dotfill & $0.04$ & $0.04$ \\ 
   Time step $\dt$ --- w/ CMC \dotfill & $0.01$ & $0.01$ \\ 
   Number of guard shells 
    \\ with the fiducial CMC \dotfill & $9$ & $9$ \\ 
   Outermost guard radius $r_{\rm max}\tablenotemark{b} $ \dotfill & $0.127$ & $0.127$ \\
   Innermost guard radius $r_{\rm min} $ \dotfill & $0.003$ & $0.003$ \\ \hline
   \multicolumn{3}{c}{Initial Disk} \\ \hline

   Toomre $Q$ \dotfill & $1.5$ & $1.5$  \\
   RMS vertical thickness \dotfill & $0.13$ & $0.05$ \\
   Truncation radius \dotfill & $5$  & $5$ \\ \hline

   \multicolumn{3}{c}{Fixed halo} \\ \hline
   $V_0$ \dotfill &  $0.7$ &  $0.7$  \\
   Core radius $c$ \dotfill & $30$ & $30$ \\ \hline

   \multicolumn{3}{c}{Accretion rule\tablenotemark{c}} \\ \hline
   Particles added per $\dt_0$ \dotfill & 80  \\
   Accretion rule \dotfill & Gaussian in $J$ \\ 
& $\bar J= 1+0.0025t$, $\sigma_J=0.5$ \\  \hline \hline 
\label{tab:one}
\end{tabular}

\tablenotetext{a}{No particles added in the weak initial bar case}
\tablenotetext{b}{See Appendix \ref{app:guard} for more details}
\tablenotetext{c}{See Sellwood \& Moore (1999) for more details}

\end{center}
\end{table}


\clearpage

\clearpage
\addtocounter{figure}{-1}

\begin{figure}[t]
\vspace{-0.2in}
\centerline{\includegraphics[angle=0, width=0.8\hsize]{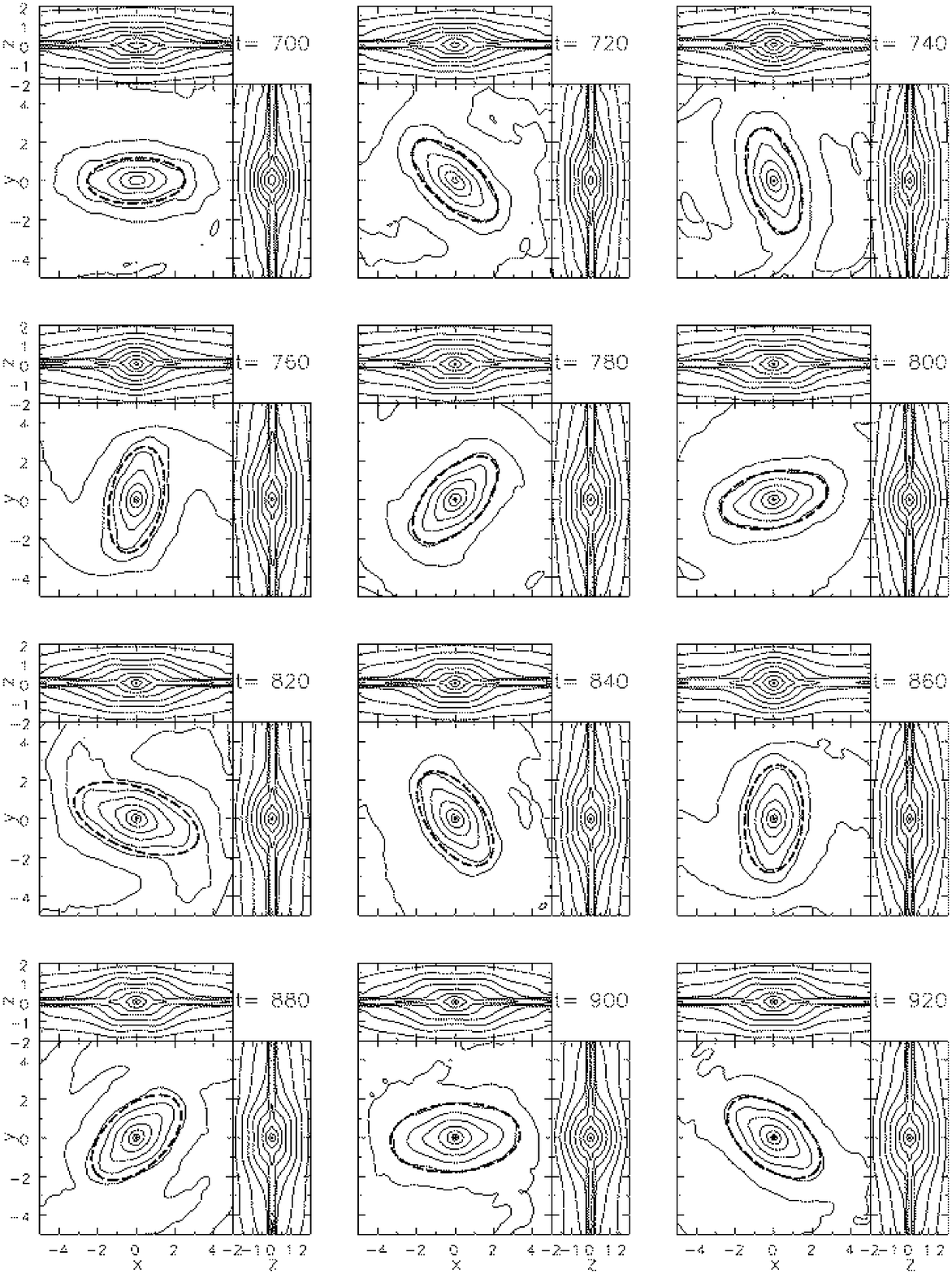}}
\caption{(b) Contours of projected density, obtained by smoothing the
particles with an adaptive kernel, at the same moments as shown in
(a).  Note that the scale is not the same.  The dashed ellipse in each
panel is the best fit ellipse with the greatest ellipticity, resulting
from the task {\tt ellipse} in IRAF. The best-fit ellipse appears to
match the neighboring density contours quite well. Contours are
separated by a constant ratio of $10^{0.4}$ (one magnitude in
projected density).}
\end{figure}

\clearpage
\begin{figure}[t]
\centerline{\includegraphics[angle=-90, width=\hsize]{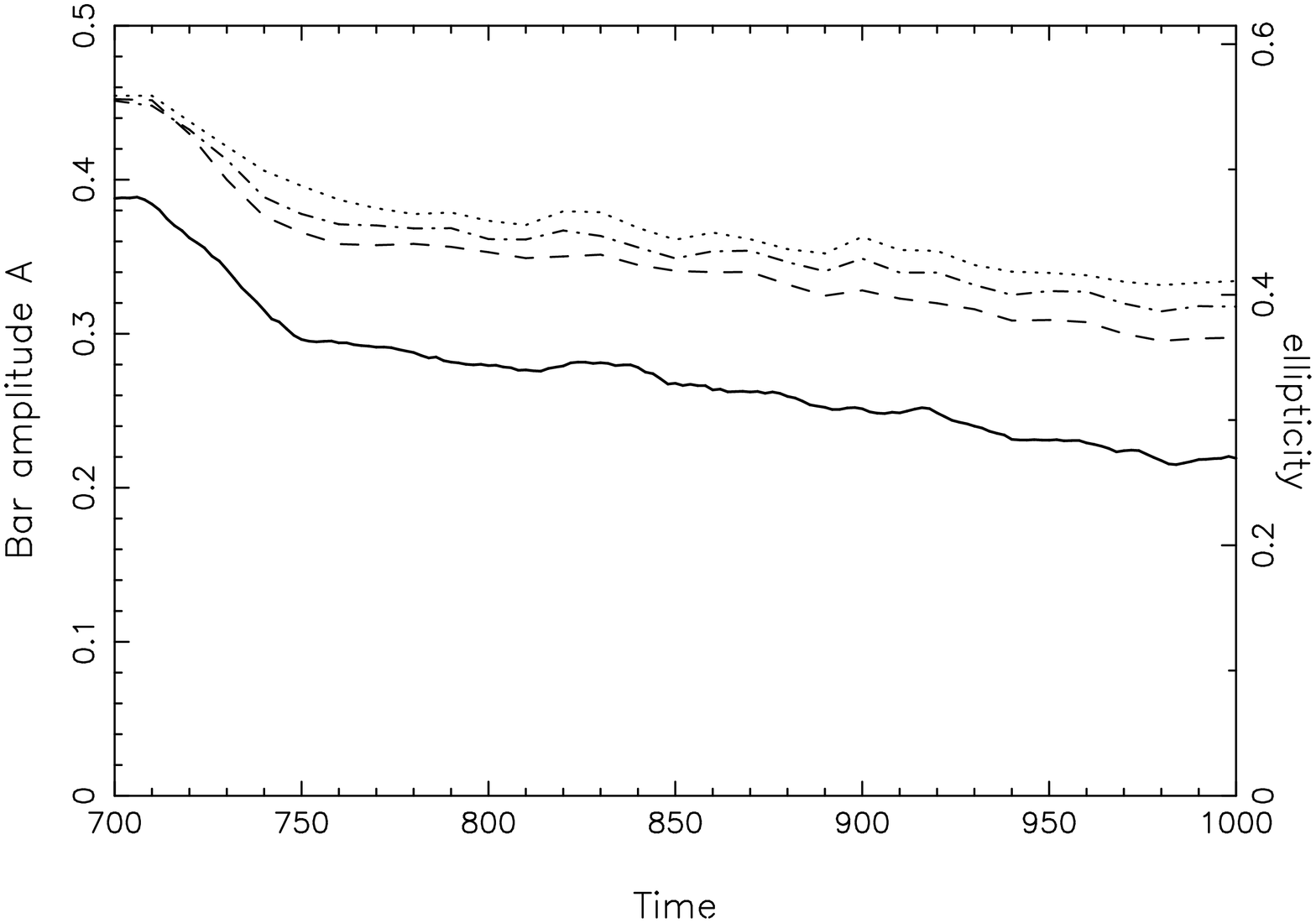}}
\caption{The time evolution of the bar amplitude $A$ (heavy solid
line) and ellipticity $e$ measured at SMA=2.0, 1.75 and 1.5 (dotted,
dash-dotted and dashed curves), respectively.}
\label{fig:ell_t}
\end{figure}

\clearpage
\begin{figure}[t]
\centerline{\includegraphics[angle=-90, width=\hsize]{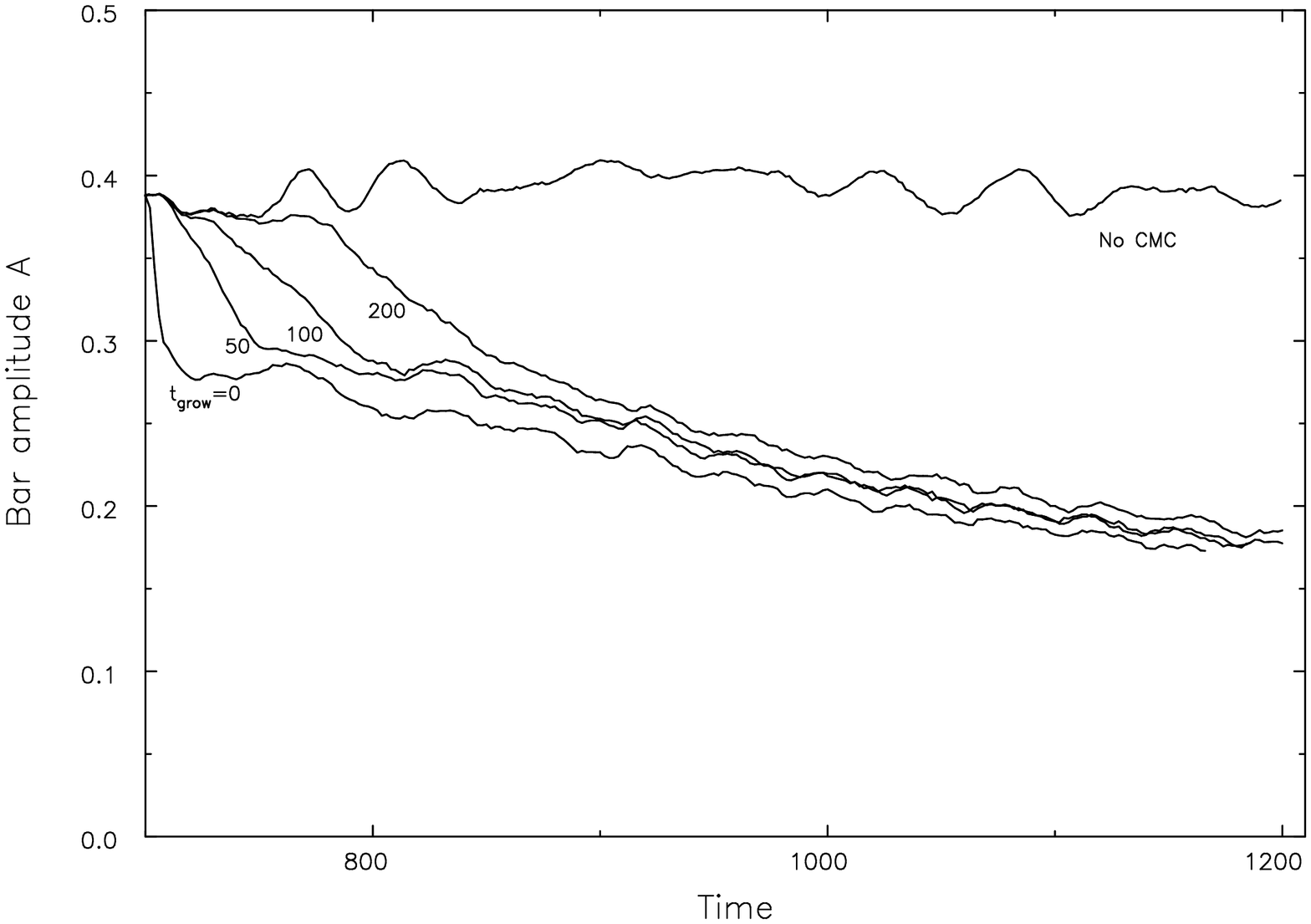}}
\caption{Evolution of bar amplitude $A$ for runs with the same CMC 
($\mcmc=0.02\mdisk$, $\ecmc=0.001$) but grown at different rates. 
The uppermost curve is a comparison run
with no CMC, the others are marked by the value of $\tg$.  The pattern
speed of this initial bar is about 50 time units.}
\label{fig:A-tgrowth}
\end{figure}

\clearpage
\begin{figure}[t]
\centerline{\includegraphics[angle=-90, width=\hsize]{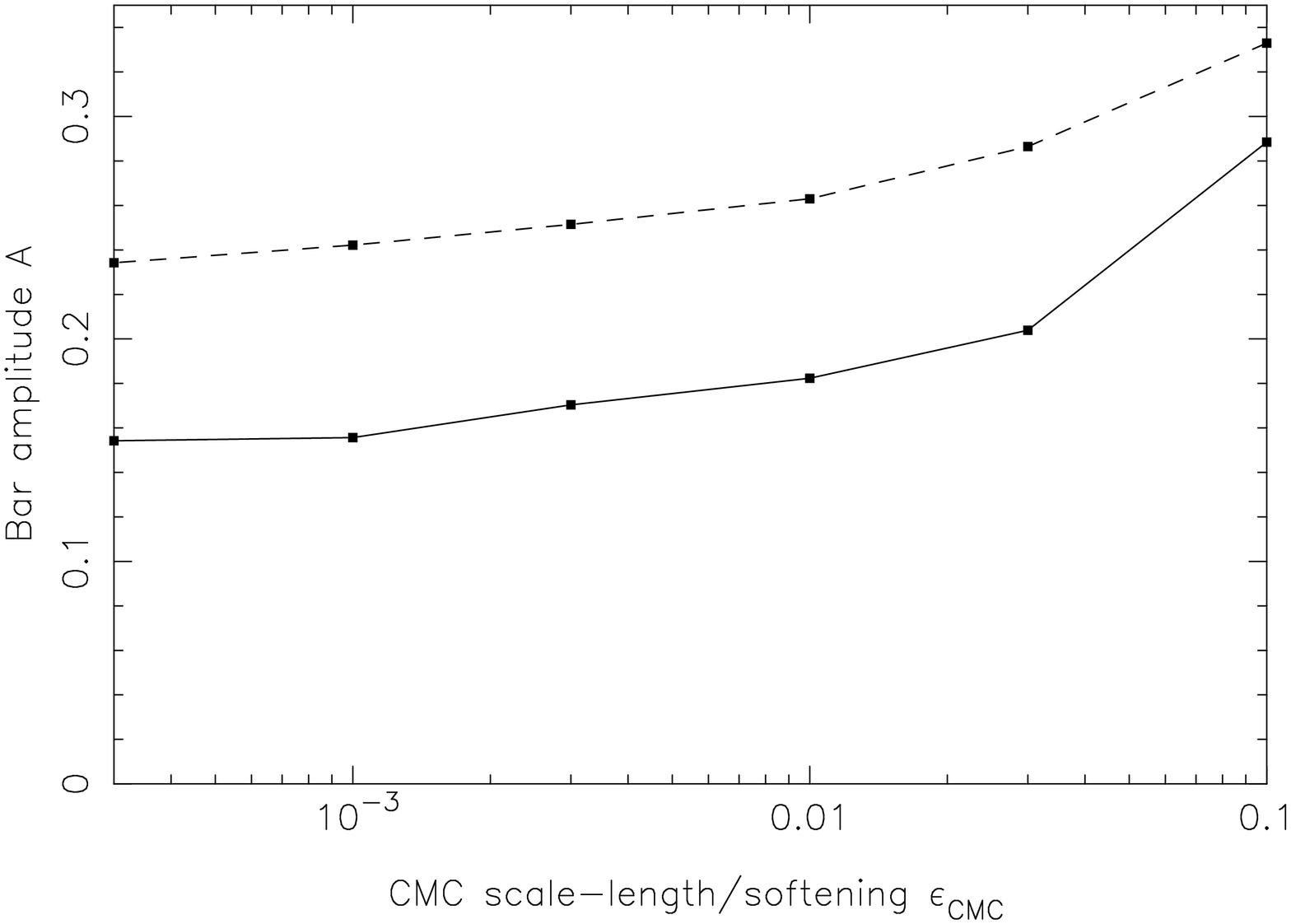}}
\caption{The bar amplitude $A$, measured 250 time units after the
central mass starts to grow, as a function of $\ecmc$, with other CMC
parameters fixed ($\mcmc=0.02\mdisk$). 
The solid and dashed curves show results for the
weak and strong initial bars, respectively.  The trend in both curves
is similar: denser CMCs cause significantly more damage to the bar,
but both curves flatten as $\ecmc$ decreases.}
\label{fig:A-ecmc}
\end{figure}

\clearpage
\begin{figure}[t]
\centerline{\includegraphics[angle=0, width=0.65\hsize]{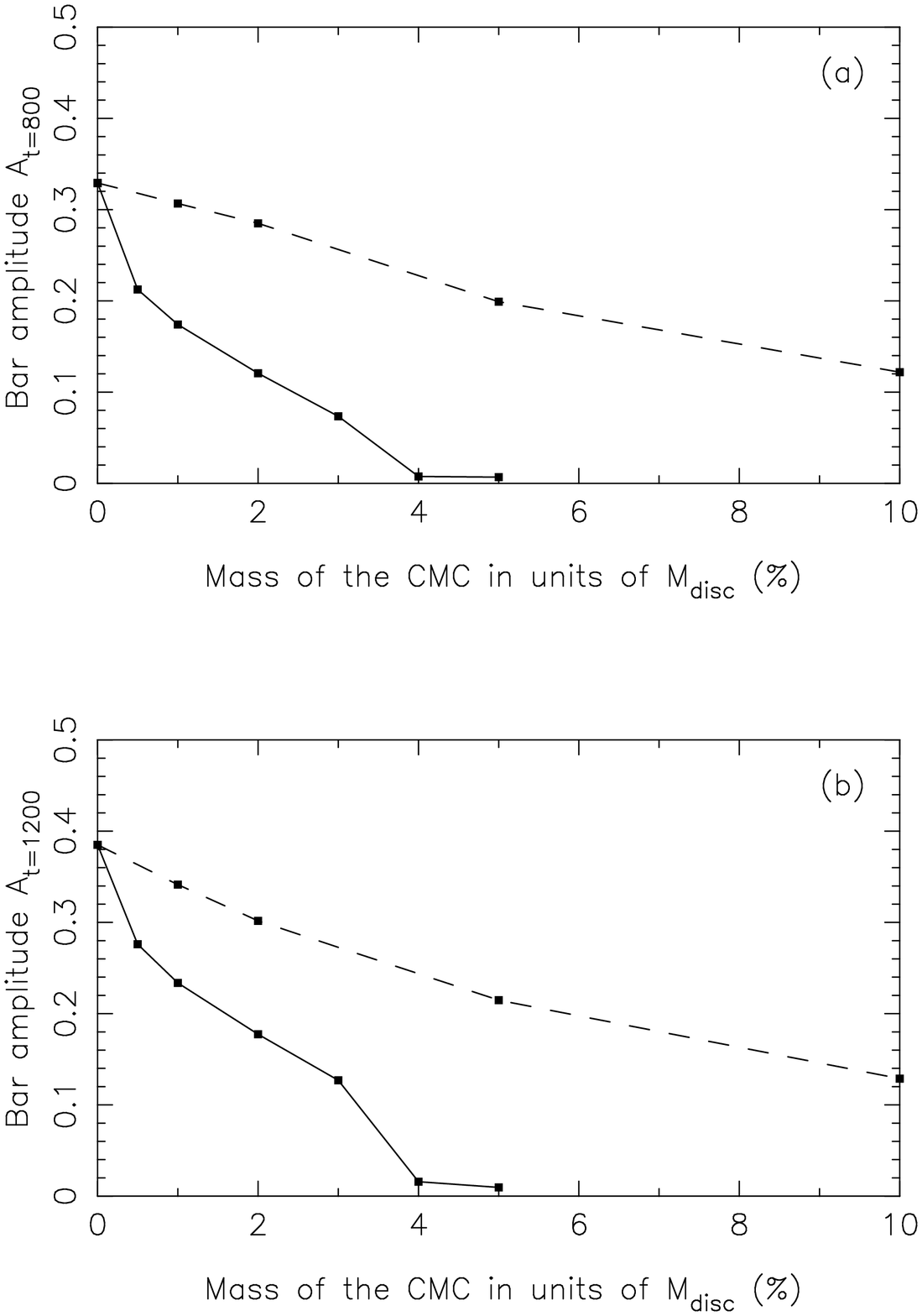}}
\caption{(a) The final amplitude of the weak initial
bar as a function of $\mcmc$, with other CMC parameters fixed. The
solid and dashed curves represent the runs with a hard ($\ecmc=0.001$)
and soft CMC ($\ecmc=0.1$), respectively.  (b) As for (a), but for the
strong initial bar. The final bar amplitude decreases continuously as
$\mcmc$ is increased, and hard CMCs always cause significantly
more damage to the bar than do soft ones.  The bar is destroyed on
short time scales by a hard CMC with a mass a few percent $\mdisk$,
but the mass of a soft CMC needs to be $\ga 0.1 \mdisk$.}
\label{fig:A-mcmc}
\end{figure}

\clearpage
\begin{figure}[t]
\centerline{\includegraphics[angle=0, width=0.5\hsize]{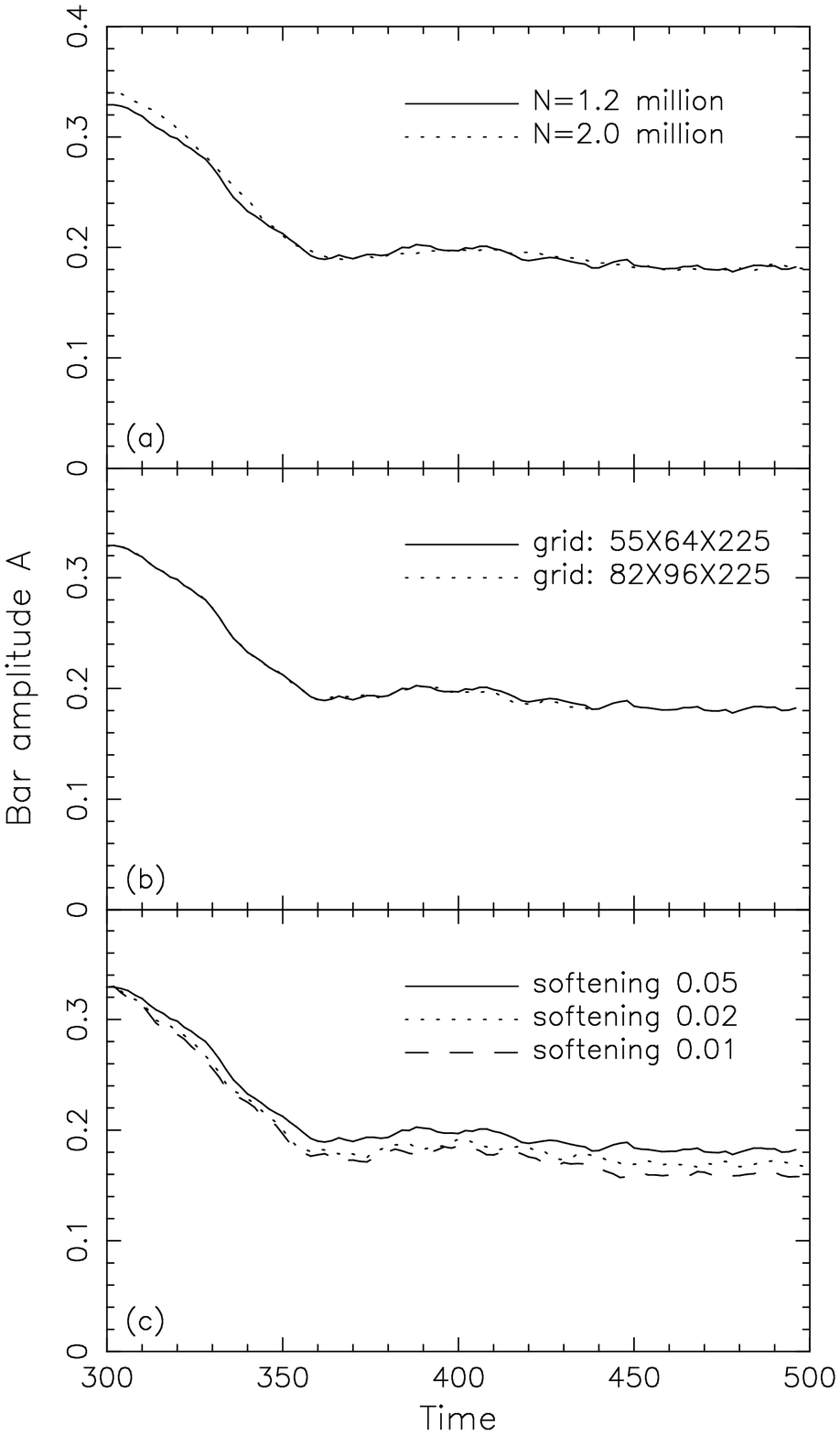}}
\caption{Tests to show that the bar-weakening behavior is little
affected when we vary the numerical parameters by a factor of two or
more around the adopted values. These tests are for the weak initial
bar, with the same CMC parameters ($\mcmc=0.02\mdisk$, $\ecmc=0.003$
and $\tg=50$).  (a) Two runs with different numbers of particles.
Note that the initial bars in these two runs start with slightly different
$A$. (b) Two with different grid sizes.  The test run with a much
finer grid yields almost exactly same result as the run with our
standard grid. (c) Tests with different particle softening from 0.05
to 0.01, with other physical parameters unchanged. }
\label{fig:basictests}
\end{figure}

\clearpage
\begin{figure}[t]
\centerline{\includegraphics[angle=-90, width=\hsize]{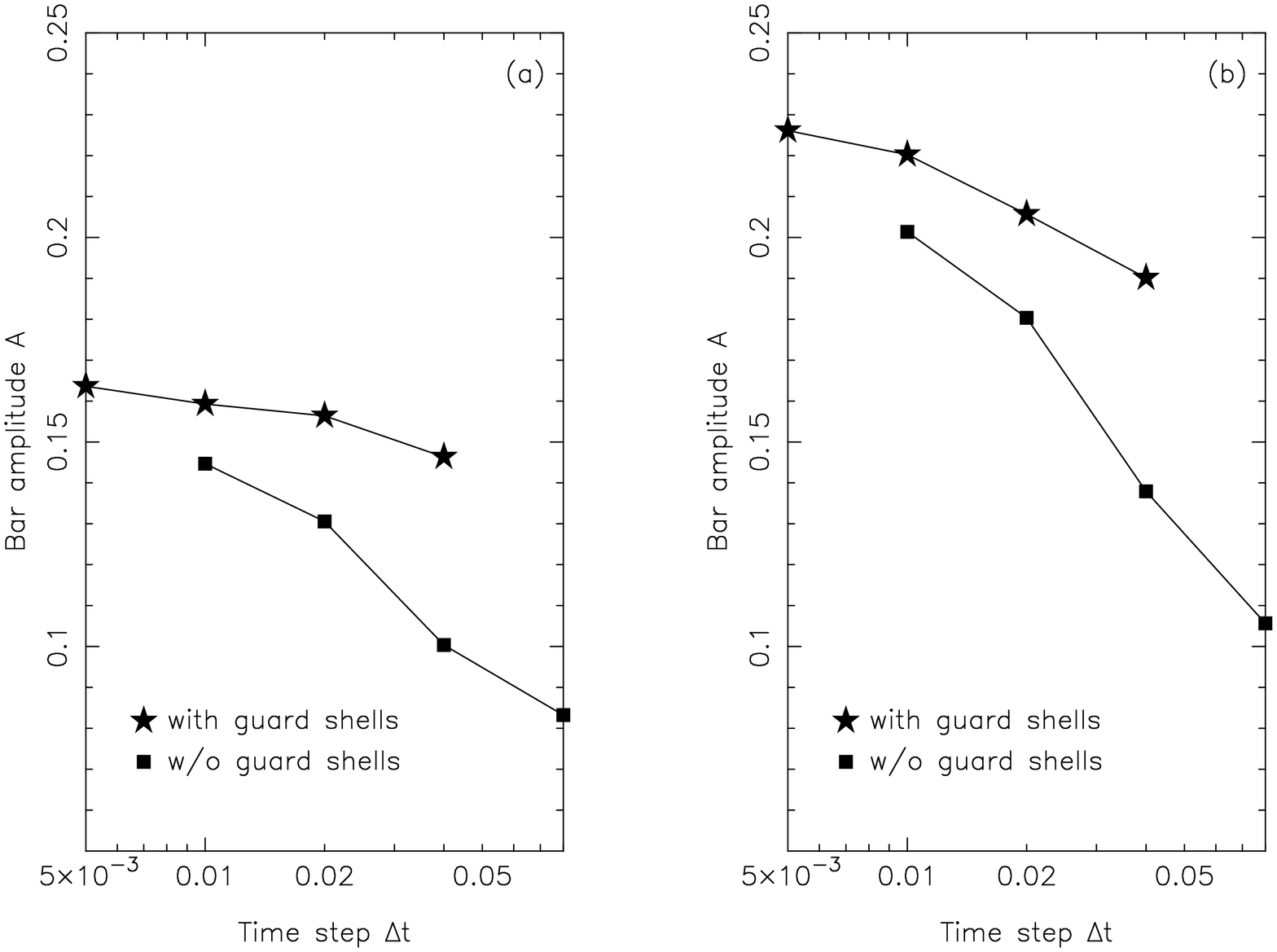}}
\caption{The bar amplitude $A$, measured at some fixed time after
$\tg$, as a function of the adopted time step $\dt$, for the weak
initial bar (a) and strong initial bar (b), with all other parameters
held fixed.  The stars and squares show, respectively, results from
runs with and without the guard shells scheme.  The final amplitude
decreases rapidly as longer time steps are employed.  Special care,
such as our guard shells scheme, is needed to obtain the correct final
amplitude when a massive hard CMC is introduced.}
\label{fig:timesteptest}
\end{figure}

\clearpage
\begin{figure}[t]
\centerline{\includegraphics[angle=-90, width=\hsize]{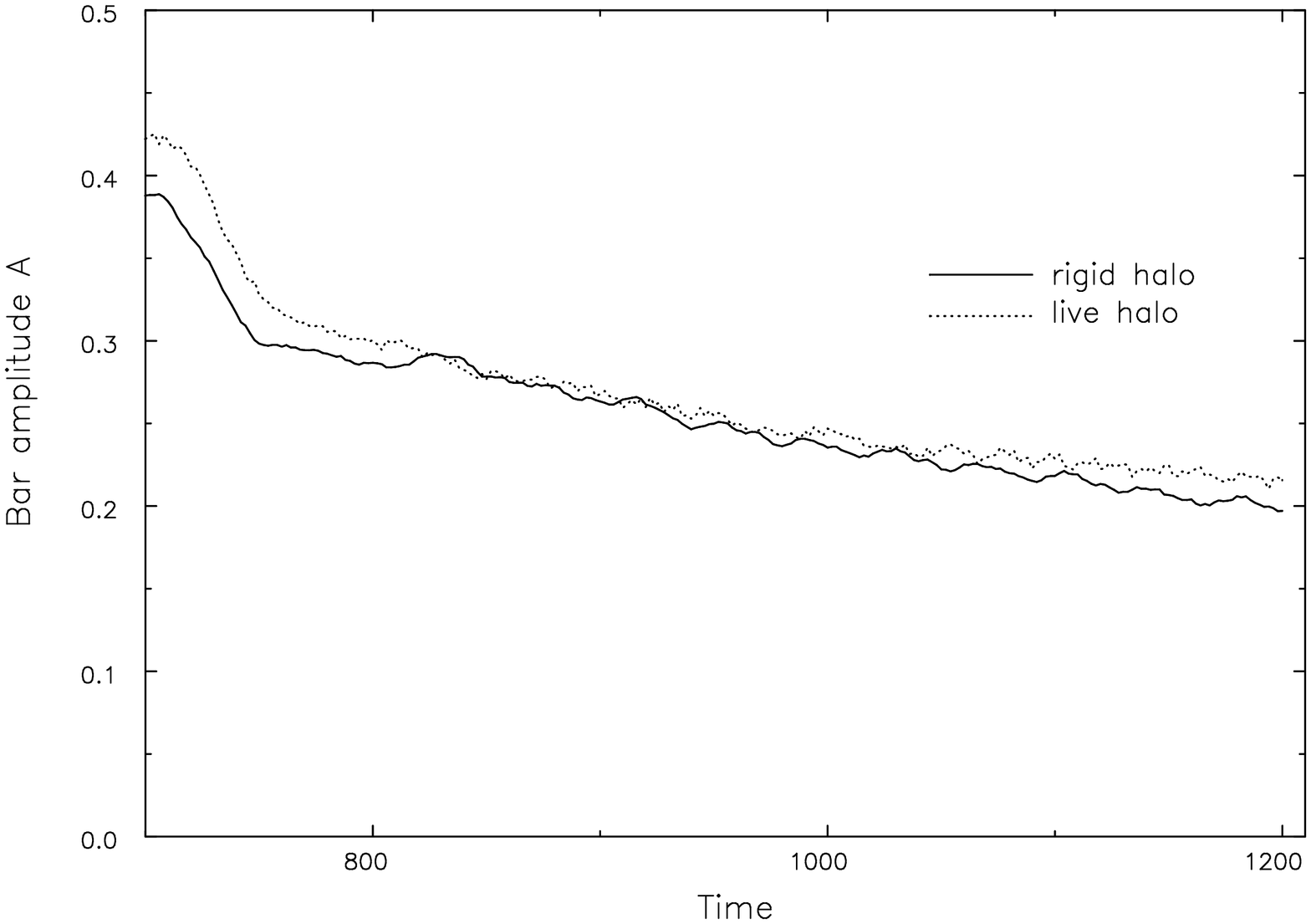}}
\caption{Comparison between runs using a rigid and a live halo for the
nearly same halo potential. The two runs have the same CMC parameters
($\mcmc=0.02\mdisk$, $\ecmc=0.003$ and $\tg=50$). 
Note that the bar amplitudes differ
slightly at the start.  The similarity the two results suggests that a
rigid halo is an adequate approximation, at least for the large-core
halo described in Eq.~(\ref{eqn:halopotential}).}
\label{fig:livehalotest}
\end{figure}

\clearpage
\begin{figure}[t]
\centerline{\includegraphics[angle=0, width=0.7\hsize]{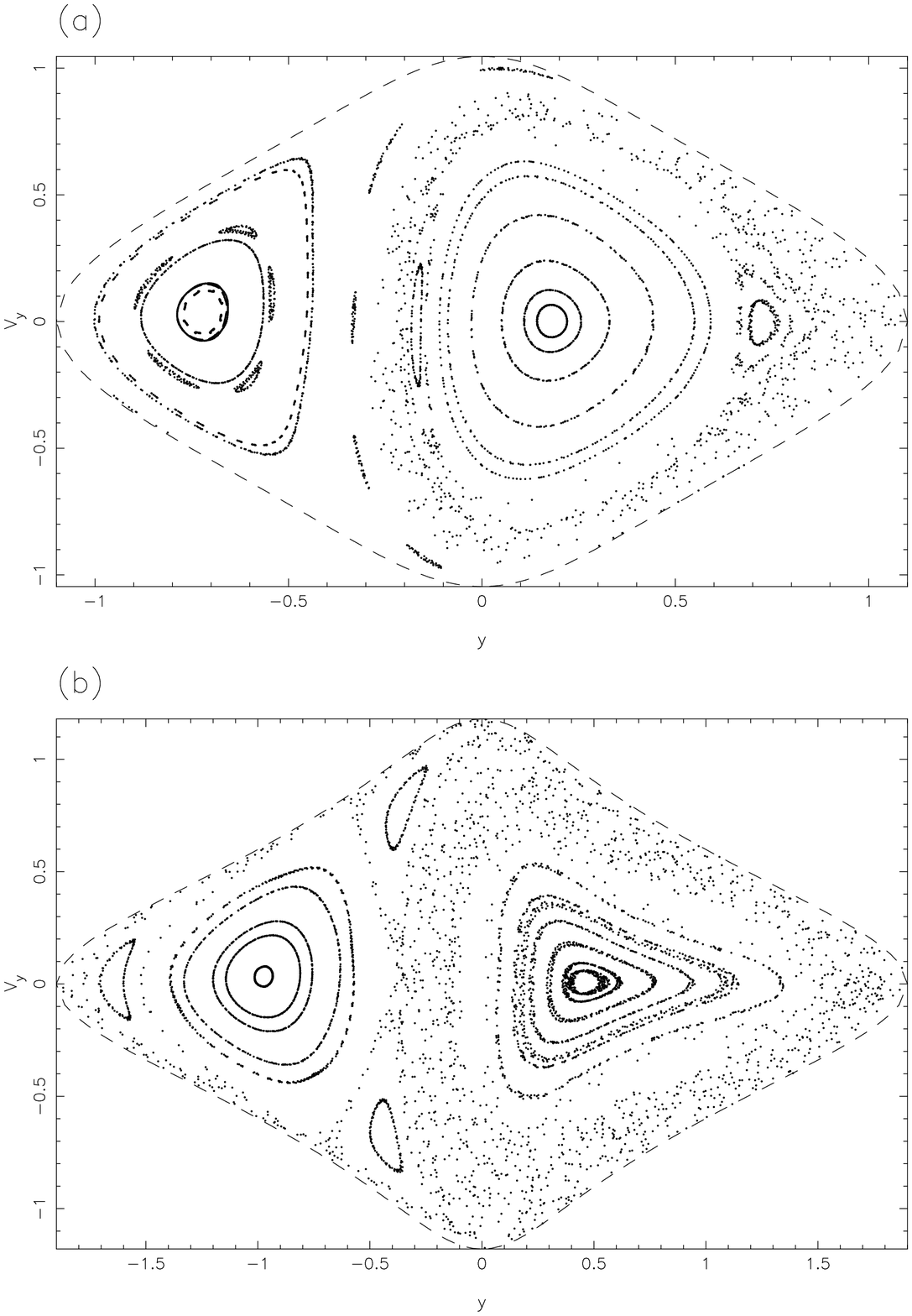}}
\caption{Surfaces of section at two energies for a case with no
central mass.  Many different test particle orbits of the chosen energy
contribute to each plot, which is constructed from the gravitational
field in the simulation at $t$=700. The dashed curve shows the
zero-velocity curve (\ie the boundary of the region energetically
accessible) for the energy specified by the limiting distance, $y_{\rm
max}$, that a particle may reach on the bar minor axis. (a) SOS for
$y_{\rm max}=1.1 \; (E_J=-0.5683)$ and (b) for $y_{\rm max} = 1.9 \;
(E_J=-0.4196)$.}
\label{fig:sos_examples}
\end{figure}

\clearpage
\begin{figure}[t]
\centerline{\includegraphics[angle=-90, width=\hsize]{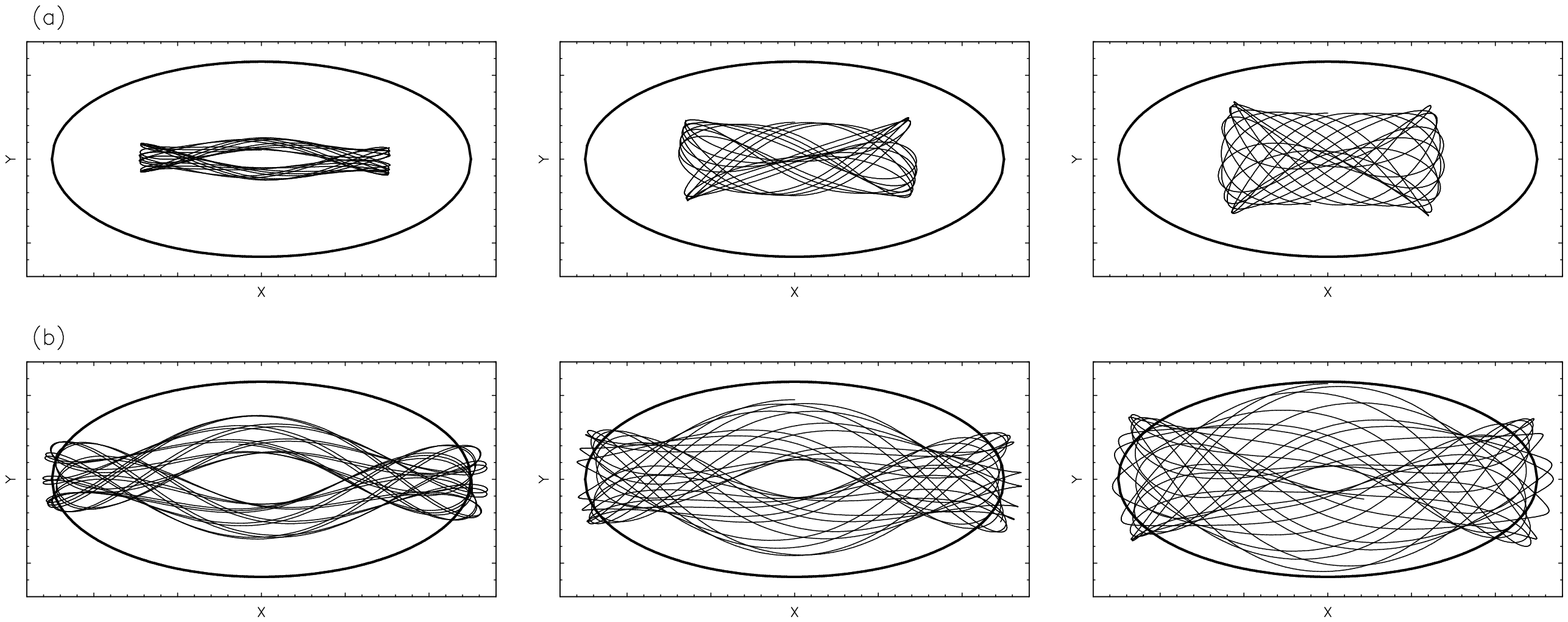}}
\caption{Sample $x_1$ orbits (in the $z=0$ plane) viewed in a frame
corotating with the bar.  The top row (a) have $E_J$ of the SOS
Figure~\ref{fig:sos_examples}(a), while the bottom row (b) have the
higher $E_J$ of Figure~\ref{fig:sos_examples}(b). The ellipse in each
panel is the bar outline for $t$=700 shown in
Figure~\ref{fig:snapshots}(b). }
\label{fig:x1_examples}
\end{figure}

\clearpage
\begin{figure}[t]
\centerline{\includegraphics[angle=0, width=0.48\hsize]{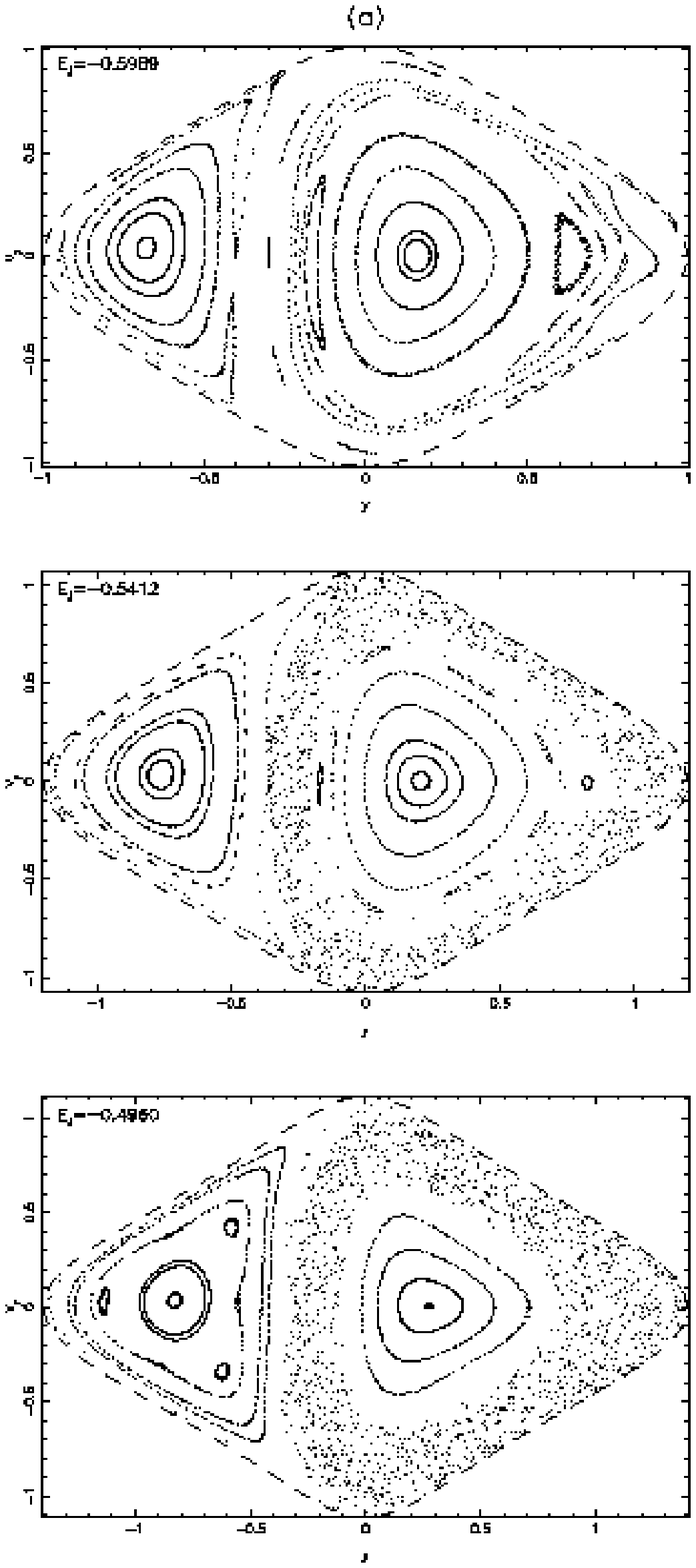}\hspace{.04\hsize}\includegraphics[angle=0, width=0.48\hsize]{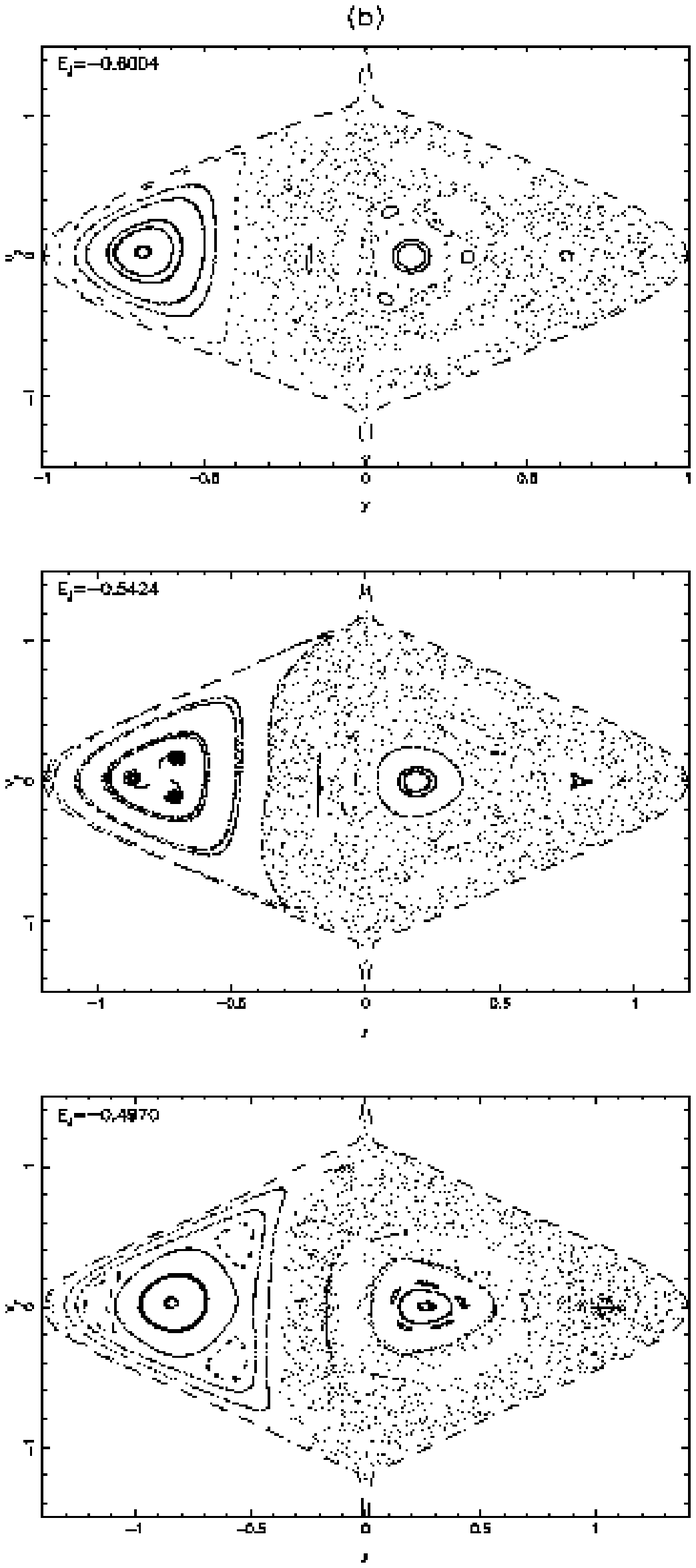}}
\caption{The evolution of SOSs for $y_{\rm max}$= 1.0, 1.2 and 1.4
(from top to bottom) after a hard CMC ($\ecmc=0.001$, $\mcmc =
0.02\mdisk$) is grown from $t=700$ to $750$. Column (a) shows the SOSs
at $t=700$, just before the CMC is added.  Columns (b) to (f) show the
SOS at the same values of $y_{\rm max}$ at times 710, 720, 730, 740 \&
750 respectively.}
\label{fig:sos_evo}
\end{figure}

\clearpage
\addtocounter{figure}{-1}
\begin{figure}[t]
\centerline{\includegraphics[angle=0, width=0.48\hsize]{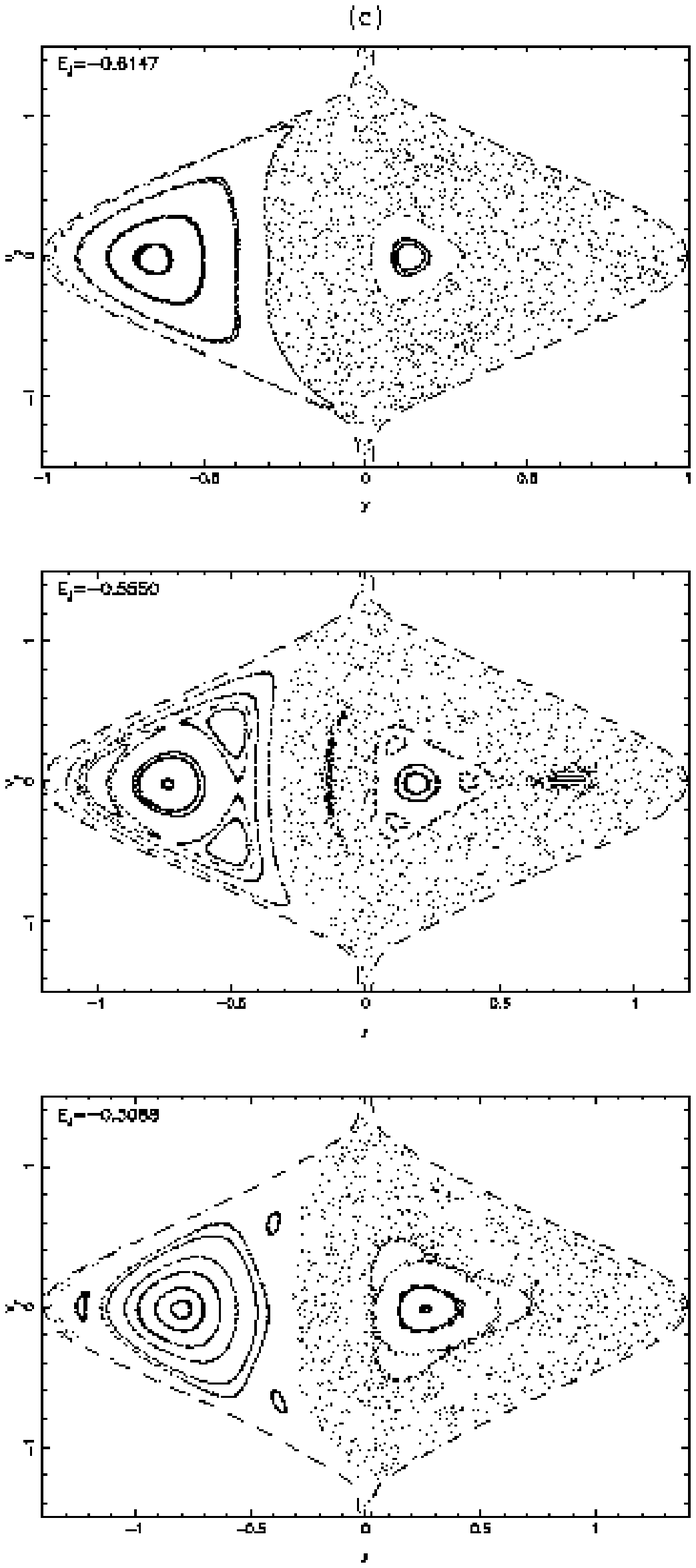}\hspace{.04\hsize}\includegraphics[angle=0, width=0.48\hsize]{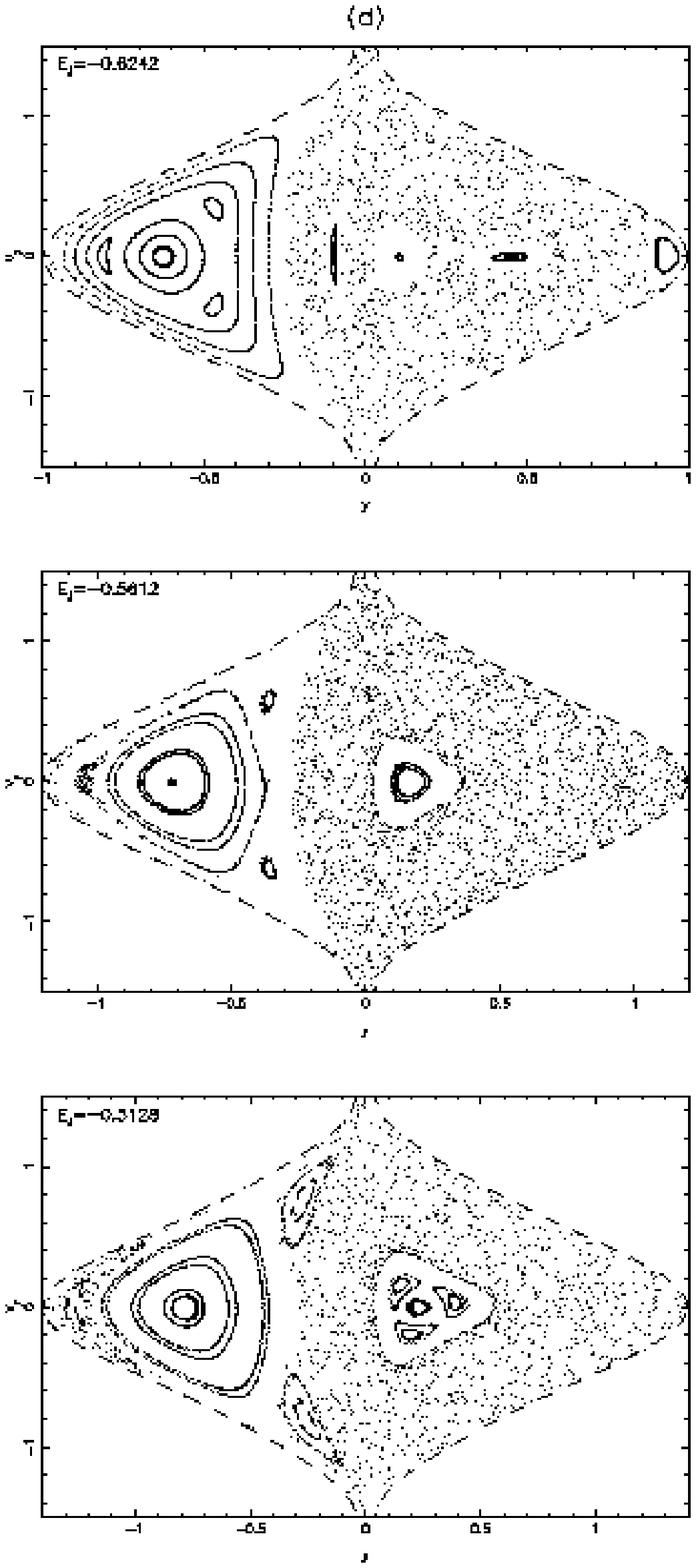}}
\caption{(c) \& (d)}
\end{figure}

\clearpage
\addtocounter{figure}{-1}
\begin{figure}[t]
\centerline{\includegraphics[angle=0, width=0.48\hsize]{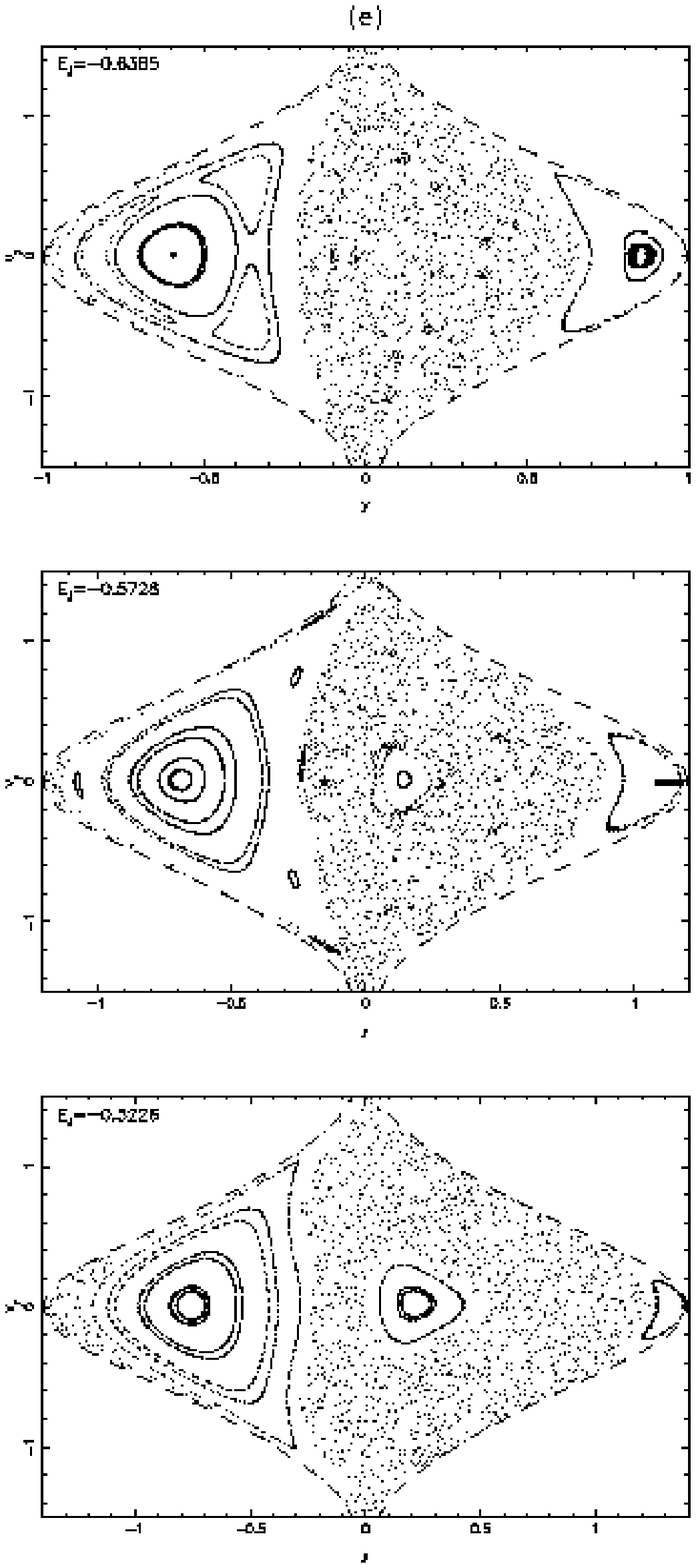}\hspace{.04\hsize}\includegraphics[angle=0, width=0.48\hsize]{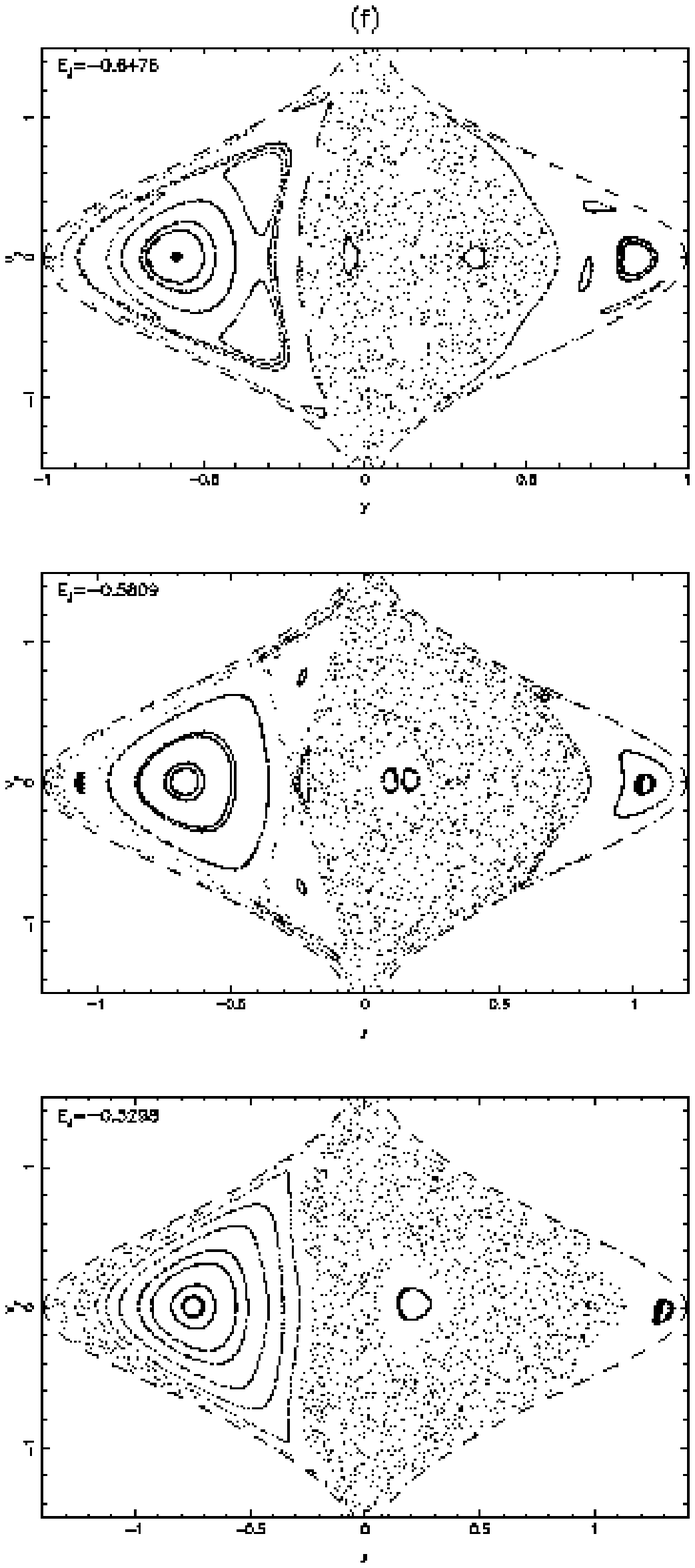}}
\caption{(e) \& (f)}
\end{figure}

\clearpage
\begin{figure}[t]
\centerline{\includegraphics[angle=-90, width=0.48\hsize]{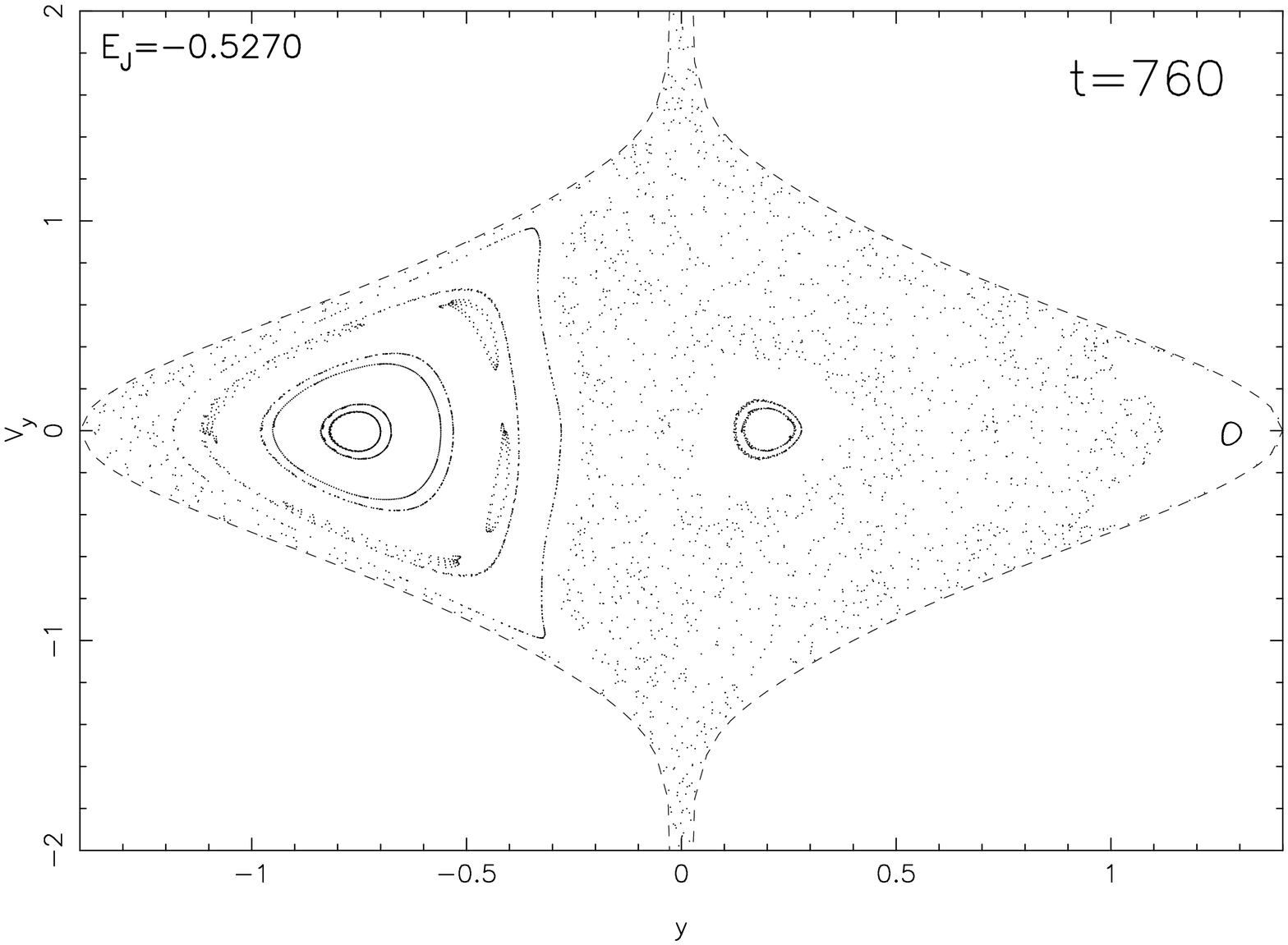}\hspace{.04\hsize}\includegraphics[angle=-90, width=0.48\hsize]{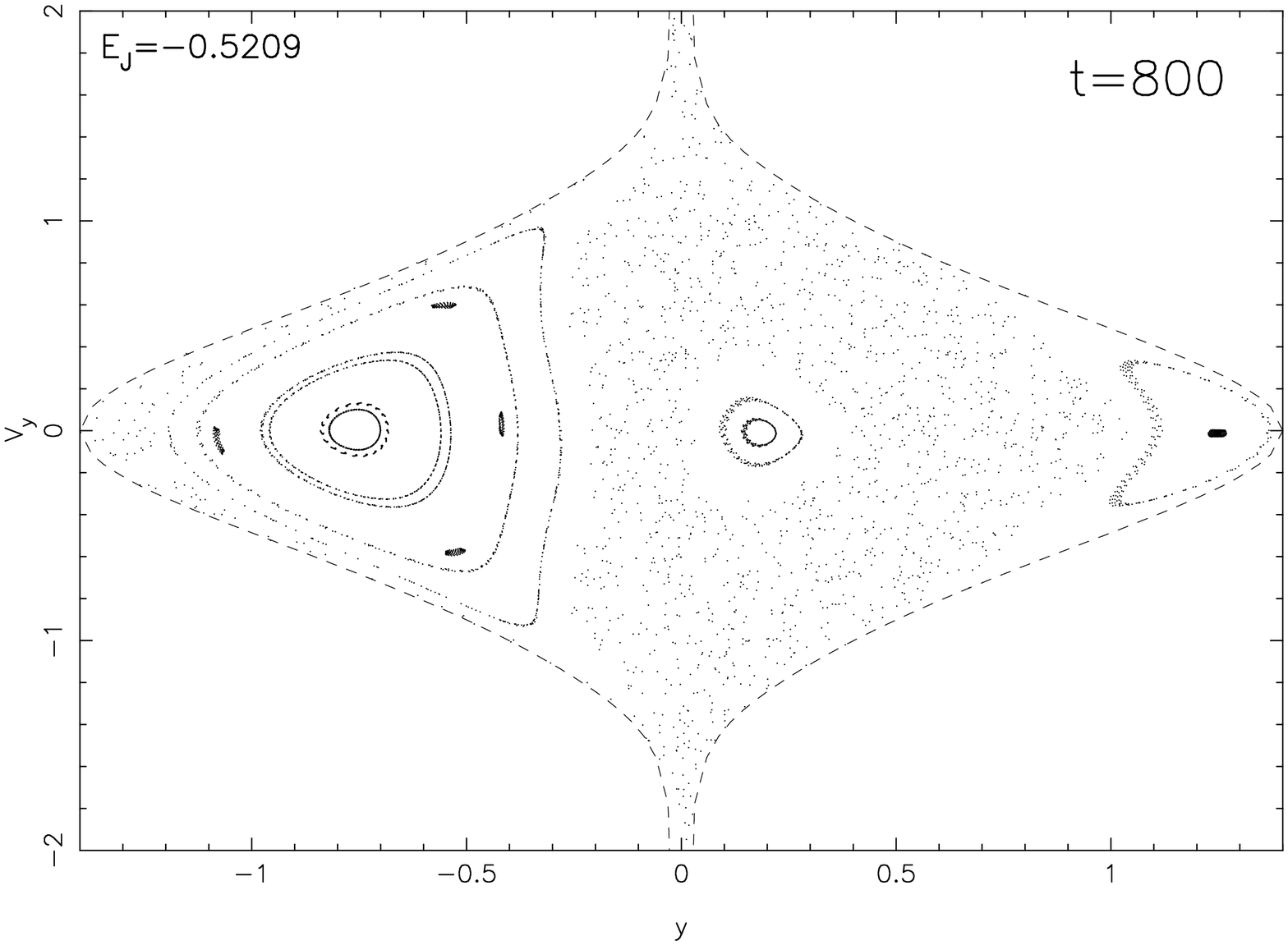}}
\centerline{\includegraphics[angle=-90, width=0.48\hsize]{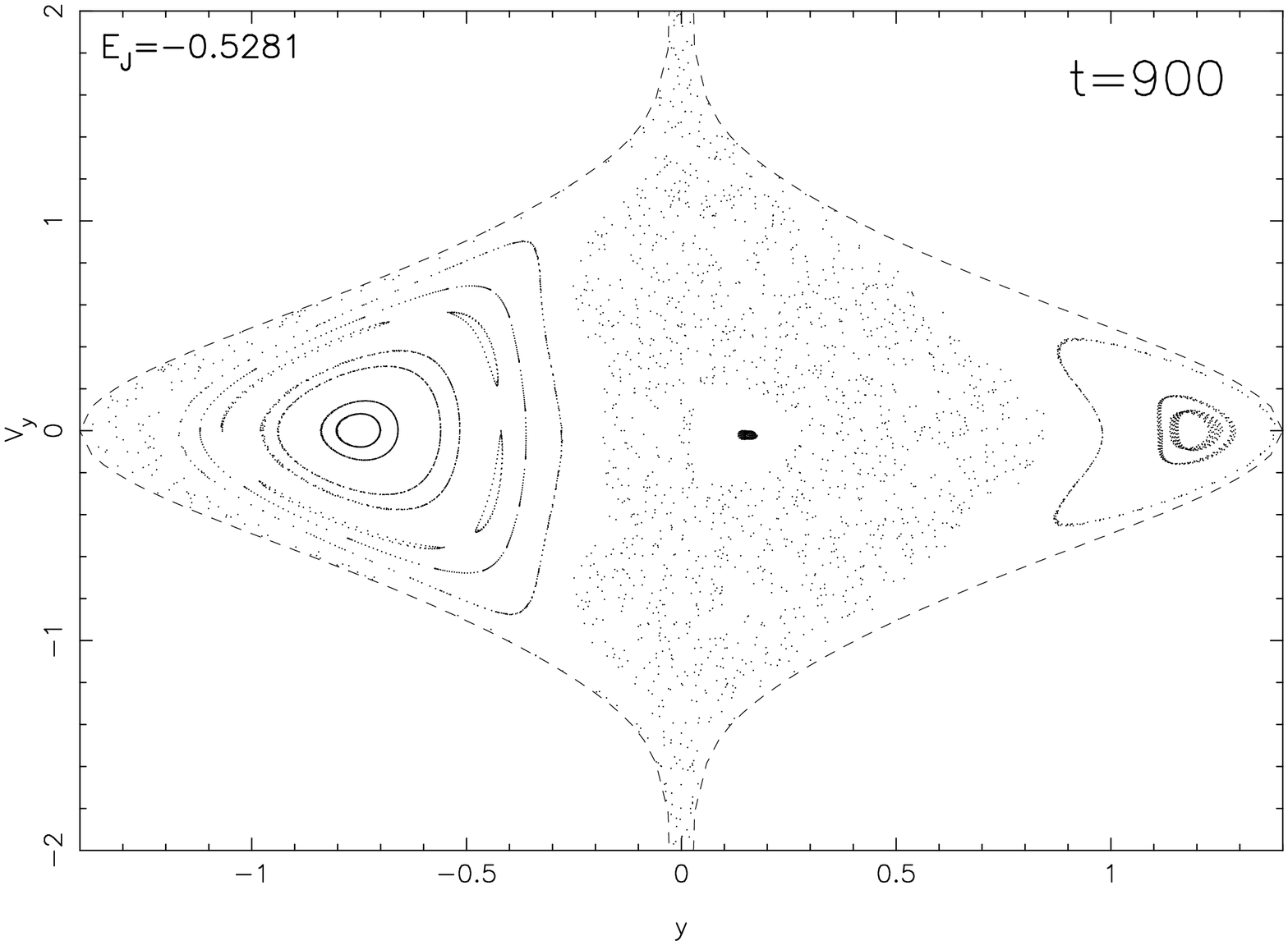}\hspace{.04\hsize}\includegraphics[angle=-90, width=0.48\hsize]{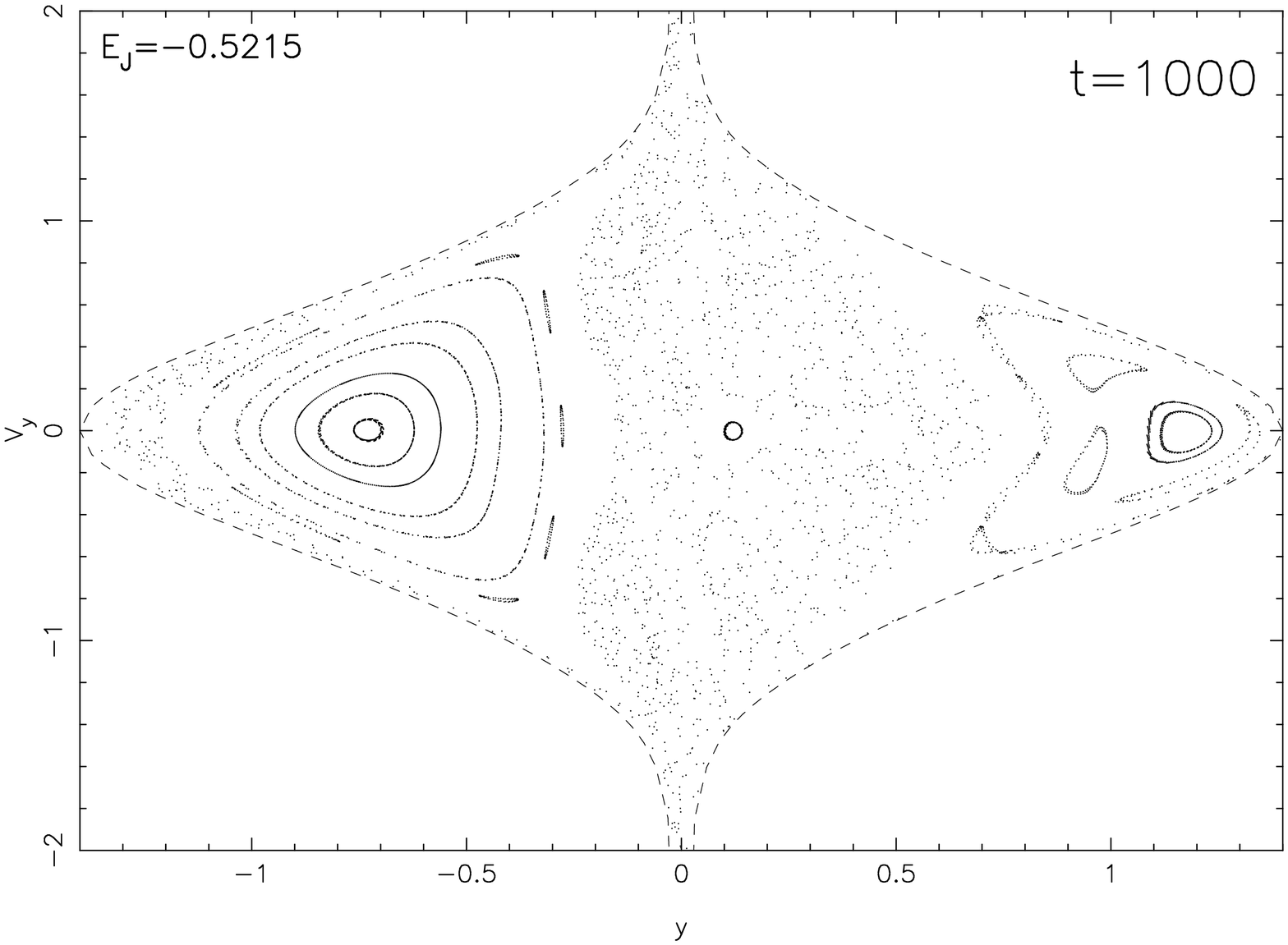}}
\centerline{\includegraphics[angle=-90, width=0.48\hsize]{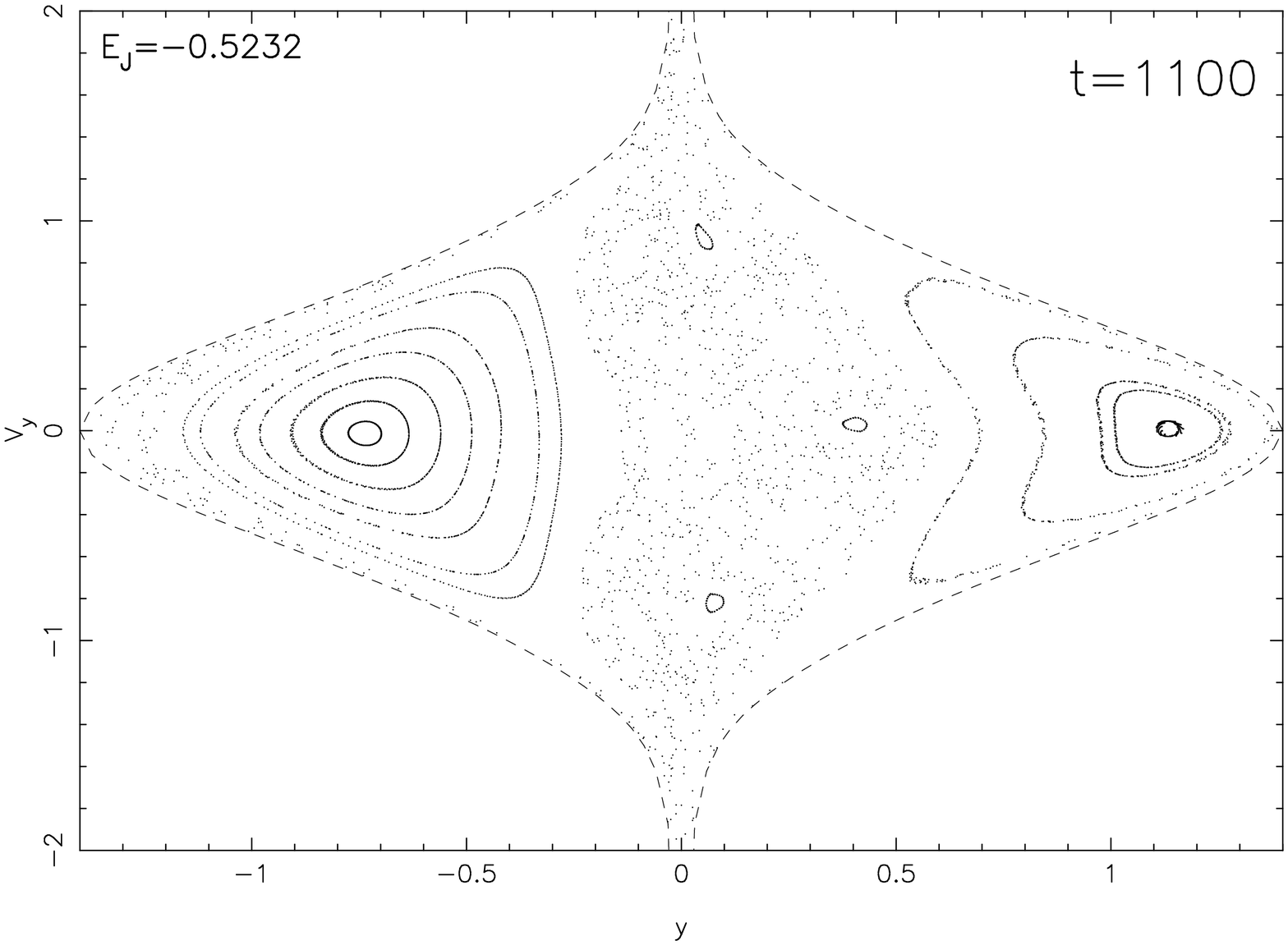}\hspace{.04\hsize}\includegraphics[angle=-90, width=0.48\hsize]{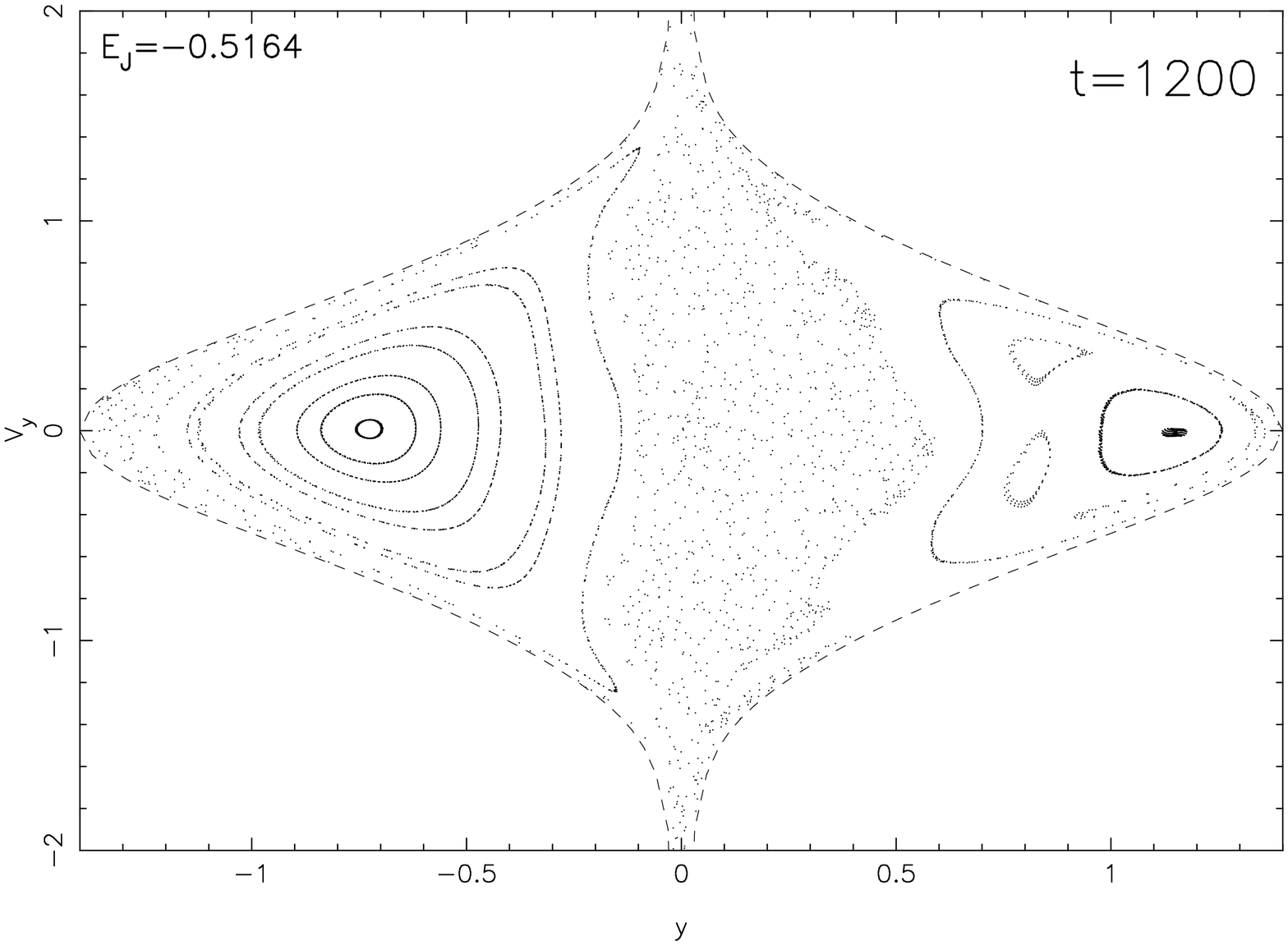}}
\caption{The time evolution of the SOS for $y_{\rm max}$= 1.4 after
the CMC has reached its final mass. The corresponding time for each
SOS is shown on the upper-right corner of each panel. The regular
region of the bar-supporting $x_1$ orbit family gradually diminishes
at this energy, consistent with the on-going decay of bar strength
after $\tg$.}
\label{fig:ymax1.4sos_evo}
\end{figure}

\clearpage
\begin{figure}[t]
\centerline{\includegraphics[angle=-90, width=\hsize]{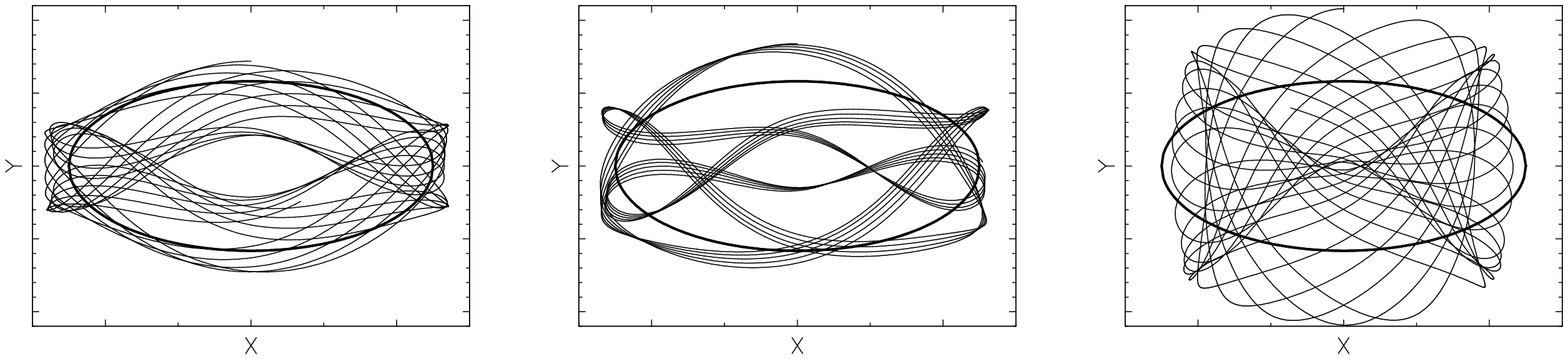}}
\caption{Three $x_1$ orbits (in the $z=0$ plane) with the same high
$E_J=-0.3874$, for the model at $t=1100$, showing that fatter $x_1$ orbits are
favored at late times. The ellipse in each panel is the outline of the initial 
bar at $t=700$.}
\label{fig:fatter_x1}
\end{figure}

\clearpage
\begin{figure}[t]
\centerline{\includegraphics[angle=0, width=.48\hsize]{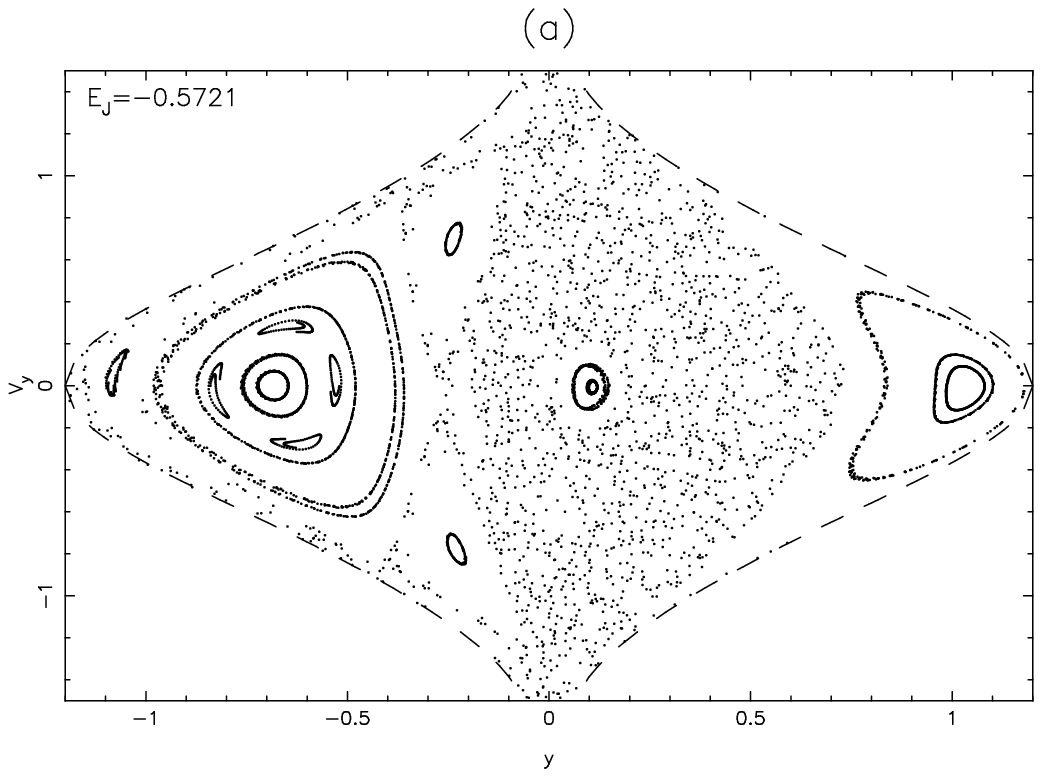}\hspace{.04\hsize}\includegraphics[angle=0, width=.48\hsize]{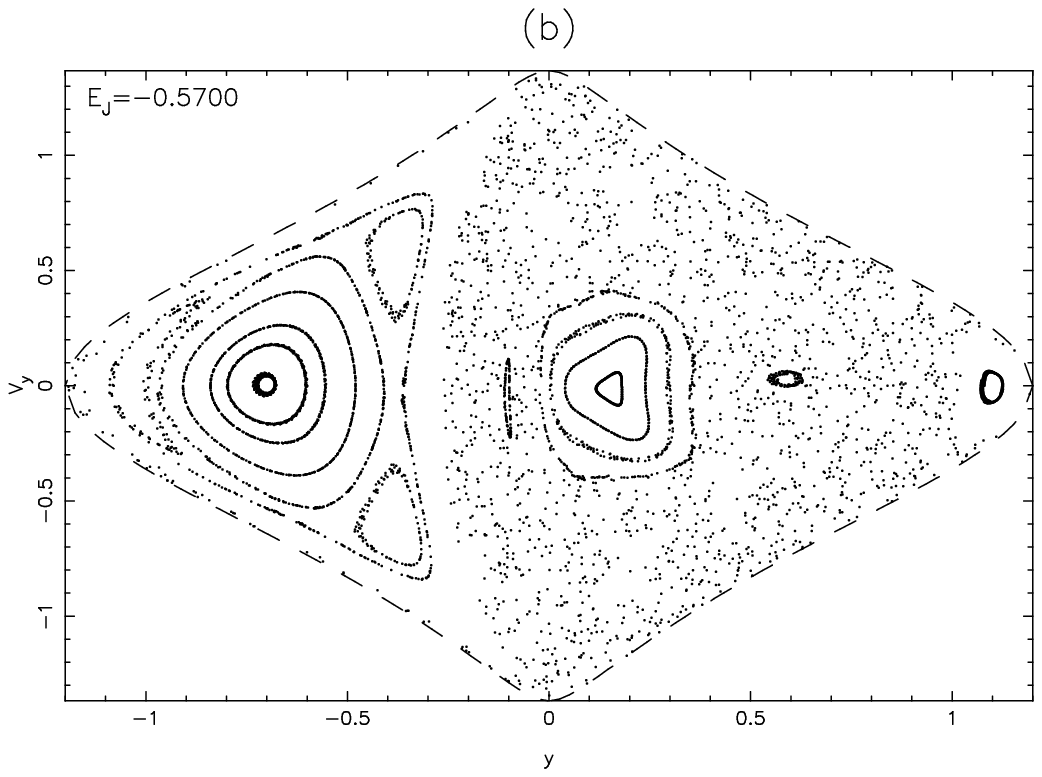}}
\caption{Comparison of SOS ($y_{\rm max}$= 1.2) at $t=800$ for the fiducial 
run with a hard CMC ($\ecmc=0.001$) on the left (a), and for a similar case
with a soft CMC ($\ecmc=0.1$) on the right (b).}
\label{fig:sos_cmc_softening}
\end{figure}

\clearpage
\begin{figure}[t]
\centerline{\includegraphics[angle=0, width=0.5\hsize]{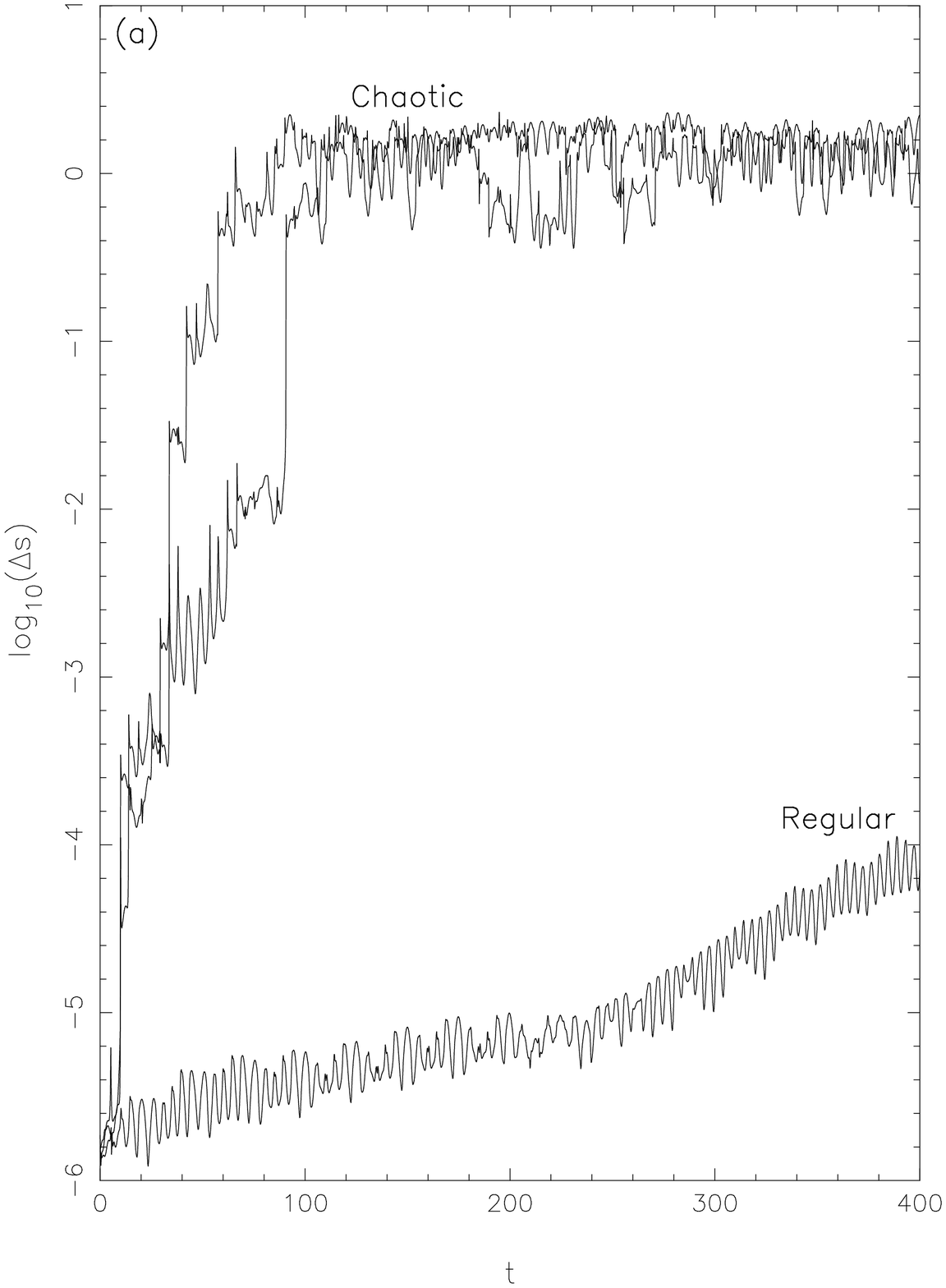}\includegraphics[angle=0, width=0.5\hsize]{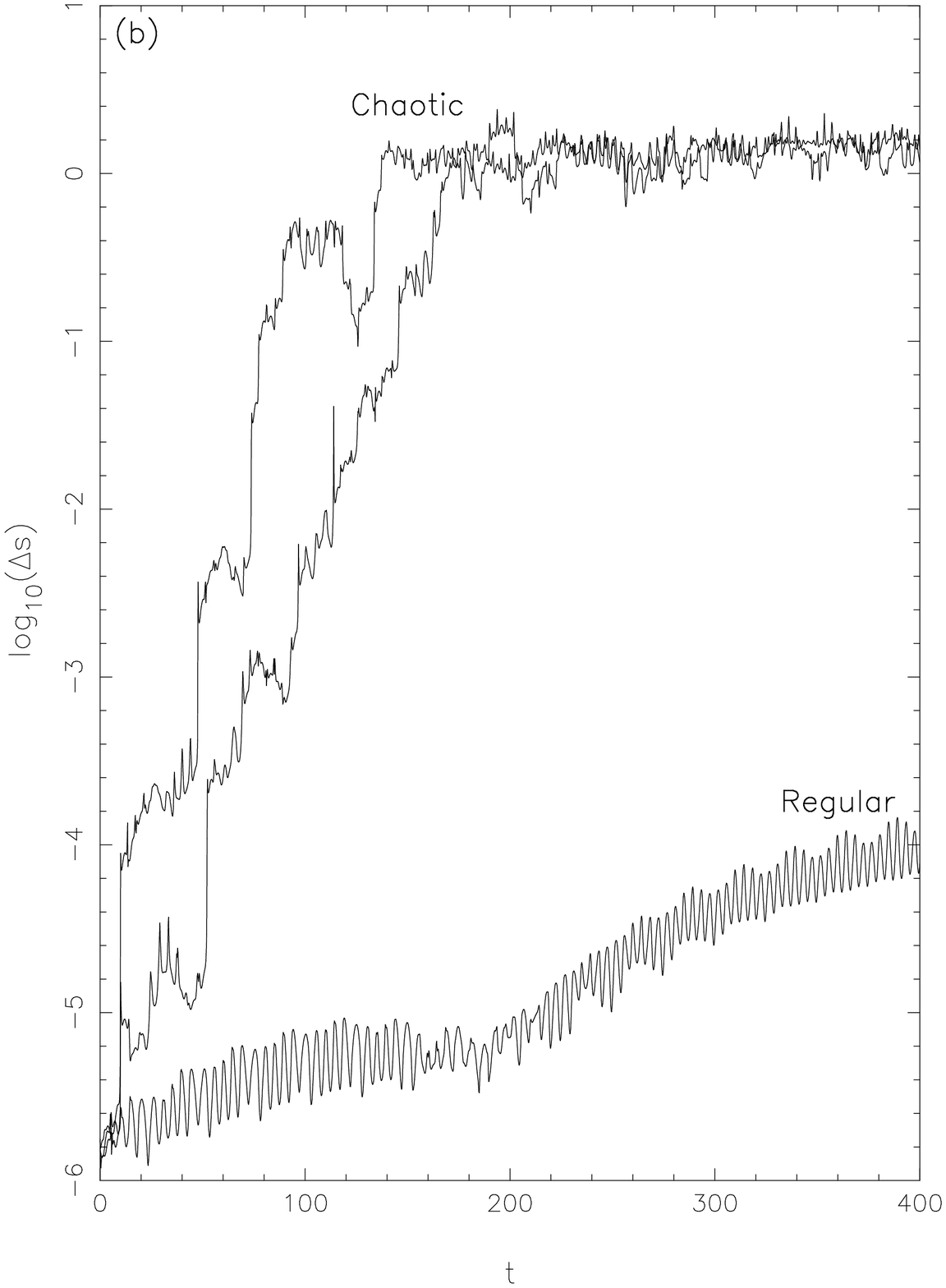}}
\caption{The time evolution ($\sim5\;$Gyr) of phase space separation due
to a infinitesimal perturbation for the potential at $t=740$,
$E_J=-0.6385$. (a) 2-D case: orbits are confined to the $z=0$
midplane. The corresponding 2-D SOS plot is the top panel of
Figure~\ref{fig:sos_evo}(e). From top to bottom for both panels, the
three curves correspond to initial $y$=0.2, 0.5 and 0.8,
respectively. (b). as for (a), but in this case full 3-D motion is
allowed.}
\label{fig:lya_ex}
\end{figure}

\clearpage
\begin{figure}[t]
\centerline{\includegraphics[angle=0, width=0.8\hsize]{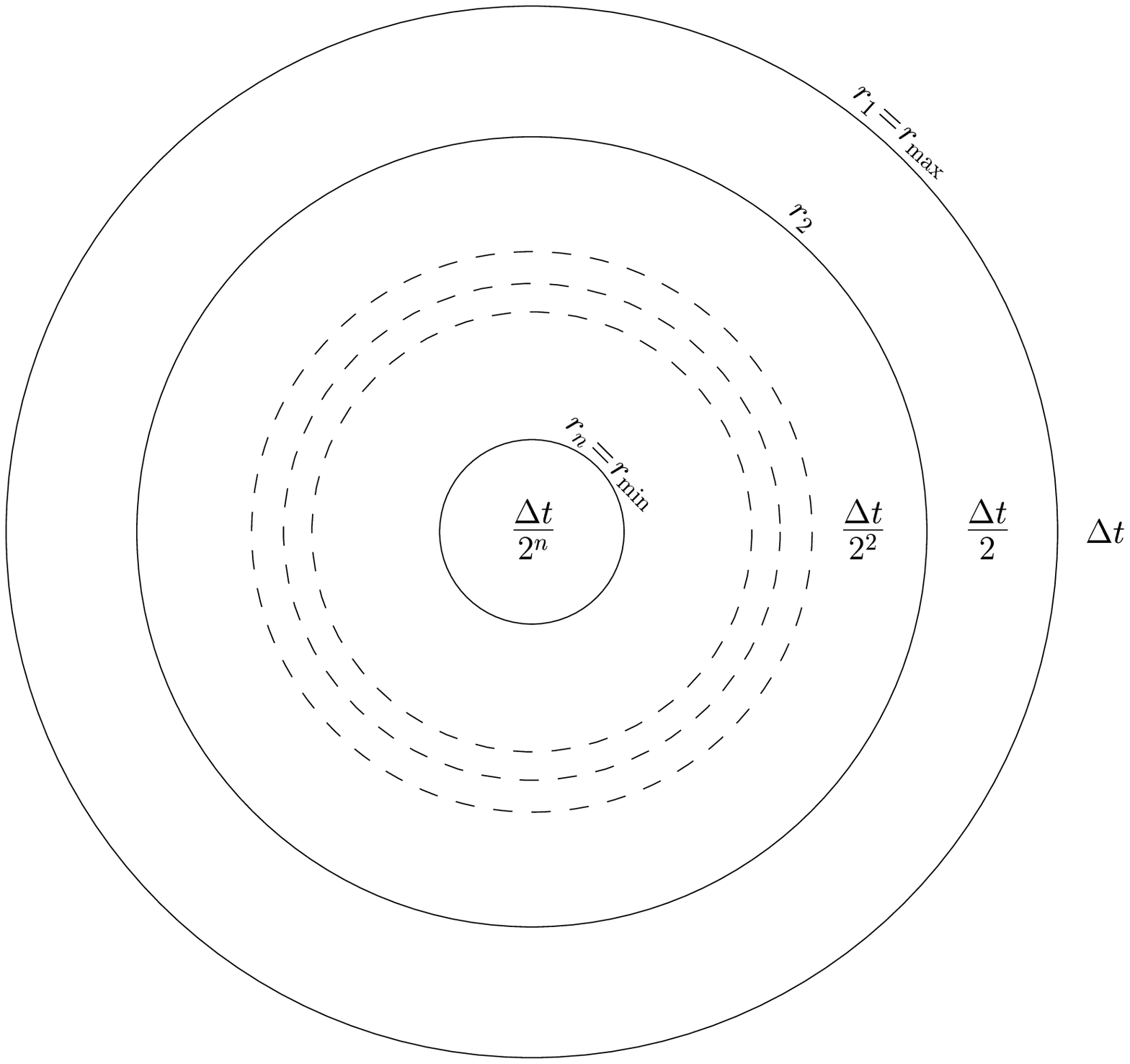}}
\caption{A sketch (not to scale) of $n$ guard shells around a compact
central mass (not shown). The ratio of the adjacent radii is
$r_{i+1}/r_i=(1/2)^{2/3}\approx 0.63$. }
\label{fig:GA}
\end{figure}

\end{document}